\documentclass[12pt]{article}

\usepackage{latexsym,amsfonts,amsmath,amsthm,amssymb,amsbsy,multirow ,textcomp,hyperref,wrapfig,datetime,verbatim,cancel,subfigure,cite}
\usepackage[dvipdfmx]{graphicx}
\usepackage[font={footnotesize,sl},bf]{caption}

\def\be{\begin{equation}}
\def\ee{\end{equation}}
\def\bea{\begin{eqnarray}}
\def\eea{\end{eqnarray}}
\newcommand{\beq}{\begin{eqnarray}}
\newcommand{\eeq}{\end{eqnarray}}

\long\def\new#1\endnew{{\bf #1}}		
\long\def\del#1\enddel{}

\def\del{\partial}

\usepackage{color}

\newcommand{\pink}[1]{\textcolor{\pink}{#1}}

\begin{document}

 \begin{titlepage}
  \thispagestyle{empty}
  \begin{flushright}
  \end{flushright}
  \bigskip
  \begin{center}
  \baselineskip=13pt {\Large \bf{(Non-adiabatic) string creation on nice slices \\ \vspace{0.3cm}  in Schwarzschild black holes}}
   \vskip1.5cm 
   \centerline{ 
   {\bf Andrea Puhm}${}^\diamondsuit{}^\odot{}^\spadesuit$, 
   {\bf Francisco Rojas}${}^\bigstar{}^\clubsuit$, 
   {\bf Tomonori Ugajin}${}^\clubsuit$
   }
  \bigskip
   \bigskip
 \centerline{\em${}^\diamondsuit$ Jefferson Physical Laboratory, Harvard University}
 \centerline{\em Cambridge, MA 02138, USA}
 \bigskip
  \centerline{\em${}^\odot$ Black Hole Initiative, Harvard University}
  \centerline{\em Cambridge, MA 02138, USA}
  \bigskip
   \centerline{\em${}^\spadesuit$ Department of Physics, University of California}
 \centerline{\em Santa Barbara, CA 93106 USA}
  \bigskip
 \centerline{\em${}^\bigstar$ Instituto de F\'{i}sica Te\'{o}rica, UNESP-Universidade Estadual Paulista}
\centerline{\em R. Dr. Bento T. Ferraz 271, Bl. II, Sao Paulo 01140-070, SP, Brazil}
   \bigskip
 \centerline{\em ${}^\clubsuit$ Kavli Institute for Theoretical Physics}
 \centerline{\em University of California, Santa Barbara, CA 93106-4030 USA}

  \vskip1cm
  \end{center}
  
  \begin{abstract}
  \noindent
  
Nice slices have played a pivotal role in the discussion of the black hole information paradox as they avoid regions of strong spacetime curvature and yet smoothly cut through the infalling matter and the outgoing Hawking radiation, thus, justifying the use of low energy field theory. To avoid information loss it has been argued recently, however, that local effective field theory has to break down at the horizon. To assess the extent of this breakdown in a UV complete framework we study string-theoretic effects on nice slices in Schwarzschild black holes. Our purpose is two-fold. First, we use nice slices to address various open questions and caveats of~\cite{Silverstein:2014yza} where it was argued that boost-enhanced non-adiabatic string-theoretic effects at the horizon could provide a dynamical mechanism for the firewall. Second, we identify two non-adiabatic effects on nice slices in Schwarzschild black holes: pair production of open strings near the horizon enhanced by the presence of the infinite 
tower of highly excited string states and a late-time non-adiabatic effect intrinsic to nice slices.

\end{abstract}

 \end{titlepage}

\setcounter{tocdepth}{2}

\tableofcontents

\section{Introduction}

A complete understanding of the physics of observers falling into black holes requires a discussion of the dynamics. Recent firewall arguments have suggested that in the presence of black holes local effective field theory may be inconsistent with unitarity of the underlying full quantum theory. Many ideas for resolving this paradox have emerged over the past few years albeit mostly at the level of effective field theory rather than using a UV completion of gravity such as string theory\footnote{Notable exceptions include the fuzzball proposal (see~\cite{Mathur:2005zp,Bena:2007kg,Balasubramanian:2008da,Skenderis:2008qn,Mathur:2008nj,Chowdhury:2010ct,Bena:2013dka} for reviews), the study of non-local effects from sting field theory~\cite{Lowe:1995ac,Polchinski:1995ta,Dodelson:2015uoa,Dodelson:2015toa} and the related work of~\cite{Silverstein:2014yza} on non-adiabatic string-theoretic effects near black hole horizons which we investigate in this paper.}. To resolve a paradox that suggests the breakdown of 
effective field theory but 
whose arguments are based on effective field theory, it seems natural that one has to go beyond the framework of effective field theory.\footnote{A different approach is 
the recently proposed soft (gravitational) hair~\cite{Hawking:2016msc} which seems to invalidate some of the assumptions made in Hawking's original calculation that led him to conclude that information is lost.} A UV completion of gravity should then capture the relevant degrees of freedom that either manifest themselves at low energies as the proposed firewall or correct the na\"{\i}ve low-energy picture.

There are several known instances, notably in extreme environments, where effective field theory does not capture the correct physics and a full analysis using a UV complete theory is required. One such example is the pair production of open strings in the presence of external electric fields~\cite{Bachas:1992bh}. Whereas in the weak-field limit one recovers Schwinger's result~\cite{Schwinger:1951nm}, for strong enough electric fields the string theory calculation shows that there is a non-adiabatic enhancement as compared to the particle result. The Hagedorn density of string states constitutes new UV physics that has to be taken into account in order to correct the effective field theory result in the strong-field regime. This is a classic example where string-theoretic non-adiabatic effects can exceed na\"{\i}ve extrapolations of effective field theory.\footnote{See also references~\cite{McAllister:2004gd,Bachlechner:2013fja,Senatore:2011sp,Amati:1987uf,Amati:1987wq,Veneziano:2004er,D'Appollonio:2013rsa,
Giddings:2006vu,Giddings:2007bw}.}

In order to identify the new degrees of freedom that may be relevant for observers crossing the horizon of black holes and to examine their effects, in particular how dramatic the departure from effective field theory is, requires studying more thoroughly the relevant aspects of the UV completion of the theory. In this note we investigate string-theoretic processes in the background of black holes and assess possible non-adiabatic effects that could trigger a breakdown of effective field theory near the horizon.\footnote{In~\cite{Bena:2015dpt} one of the authors studied whether the formation of a horizon could be avoided altogether through quantum tunneling of matter (branes) into horizon-size microstate geometries. In this note we study another tunneling process, namely that of pair creation of open strings in the fixed black hole geometry.}

It turns out that the above example of open string pair production in external fields lends itself to study certain non-adiabatic effects in black hole backgrounds. In~\cite{Bachas:1995kx} Bachas has shown that the process of creating pairs of open strings in an external field has a T-dual description in terms of open strings created between pairs of D-branes that undergo a large relative boost. Incidentally, large boosts occur naturally in the background of black holes when trajectories approach the horizon and one may wonder whether this can lead to non-adiabatic effects for infalling observers? 

Indeed, in~\cite{Silverstein:2014yza} Silverstein has argued that large boosts may provide a dynamical mechanism for firewalls catalyzed by late-time infalling observers. In a dynamical thought experiment one imagines open strings stretched between D-particles moving on infalling trajectories in the black hole geometry. At late times at least one of the D-particles is to fall into the black hole thus playing the role of a late-time infalling observer. To address the question of whether an infalling observer sees a departure from effective field theory upon crossing the horizon, in this thought experiment, one computes the open string pair production rate. A non-adiabatically enhanced rate as compared to the na\"{\i}ve field theory expectation could enable an infalling observer to detect a departure from effective field theory and perhaps even encounter a firewall.

The intuitive picture of strings extended between D-branes moving on trajectories in the black hole geometry suggests a growing mass as they are stretched apart that can lead to a non-adiabatic production. Estimating the rate of string production appears rather difficult at first sight (the growth of proper length of the string na\"{\i}vely depends on the slicing which cannot affect the physics). There exist, however, approaches that nevertheless allow to reliably estimate the level of production rate such as the semi-classical worldsheet path integral method developed in~\cite{Silverstein:2014yza,Joeunpublishednote}. This method has been tested in the near-horizon region of black holes which are described by Rindler space (outside) and Milne space (inside), reproducing the flat space results of~\cite{Bachas:1992bh,Bachas:1995kx}. In~\cite{Silverstein:2014yza} it was furthermore shown that using these methods to compute open string pair production between 
pairs of D-branes on relatively 
relativistically boosted trajectories in the Schwarzschild geometry can 
yield 
a non-adiabatically enhanced production at the horizon; this effect was proposed as a dynamical mechanism for firewalls.

\bigskip
There are several issues that could affect (the interpretation of) the results of~\cite{Silverstein:2014yza} which therefore deserve further attention. 

Applying the above method to compute open string pair production to the entire Schwarzschild geometry is problematic. Schwarzschild coordinates do not capture the entire spacetime and, moreover, are not smooth across the horizon. Therefore, in~\cite{Silverstein:2014yza} the analysis was carried out in a Gullstrand-Painlev\'{e} slicing of the black hole geometry; these coordinates are smooth across the horizon but eventually intersect the singularity. One may worry whether the string action contains contributions from the singularity and if so whether these contributions could be responsible for the non-adiabatically enhanced string production rate?

To avoid this potential problem, in the present paper, we investigate string production in a ``nice'' slicing of the Schwarzschild geometry: nice slice coordinates are smooth across the horizon \emph{and} avoid the singularity. The notion of nice slices was first explicitly discussed in~\cite{Lowe:1995ac,Polchinski:1995ta}\footnote{The authors comment that the existence of such a set of surfaces is implicitly assumed in much of the existing literature on black hole evaporation even if seldom spelled out. They cite their private communication with R.~M.~Wald at the Santa Barbara ITP conference “Quantum Aspects of Black Holes” in June 1993 as the first explicit construction of nice slices.} in the context of the information loss problem to circumvent several difficulties coming from potential quantum gravity effects in black holes. Nice slices avoid regions of strong spacetime curvature and yet smoothly cut through the infalling matter and the outgoing Hawking radiation so that both sets of particles have low 
energy in the local frame of the slice and the use of effective field theory to describe the evaporation process is (at least formally) justified. 
Since the nice slicing of the black hole geometry was devised to provide a well-defined semi-classical description of the near horizon physics, it would be surprising if we could see non-adiabatic effects, let alone a firewall. Hence, besides avoiding a potential contribution from the curvature singularity in the string production rate, we can also test the validity of effective field theory on nice slices. This provides a separate motivation for the current work.
Moreover, in contrast to the Gullstrand-Painlev\'{e} slicing, nice slices naturally introduce a well-defined cutoff surface, the last nice slice. Hence, while we cannot define an S-matrix for infalling observers, there may nevertheless be a way to define some sort of asymptotic {\it in} state (the infinite past far away from the black hole) {\it and} asymptotic {\it out} state (the infinite future at the last nice slice inside the black hole).

To determine whether the non-adiabatic effects found in~\cite{Silverstein:2014yza} are caused by the fact that the Gullstrand-Painlev\'{e} coordinates are not globally well-defined we apply the first quantized method for computing string pair production to a nice slicing of the black hole geometry. We find a general expression for the Bogoliubov coefficient for a given {\it in} and {\it out} slice in terms of the boost and impact parameter between the D-branes and their energy per rest mass {\it via} the WKB approximation. 
We identify the regime of parameters where the rate of open string pair creation is enhanced. In the limit of both large and small ratio of energy per unit mass of the infalling D-branes, strings are non-adiabatically produced.
To better understand the implications of this result we repeat the string production calculation in a nice slicing of the near-horizon region which is given by Rindler space. As in the flat space result of~\cite{Bachas:1995kx} we find enhanced production for a large relative boost between the D-branes.

Although pair production rates are only defined globally, in~\cite{Silverstein:2014yza} a local quantity, denoted by $\omega^2(T)/\dot{\omega}(T)$, was introduced which estimates the magnitude of pair production at a fixed time $T$. Based on this figure of merit, it was argued that there is a large non-adiabatic string production rate when the second D-brane (serving as the late-time infalling observer) reaches the horizon, and that this rate continues to grow inside the horizon towards the singularity. Since this quantity is not covariantly defined one may worry that the conclusion one draws from it depends heavily on the observer and the slicing one uses. 

To address this issue we compute the figure of merit on nice slices and compare it to the expression obtained in a Gullstrand-Painlev\'{e} slicing. Interestingly, its evaluation at the horizon in a nice slicing of both, the entire black hole geometry and its near-horizon region which is given by Rindler space, yields an expression that behaves qualitatively like the one obtained in a Gullstrand-Painlev\'{e} slicing~\cite{Silverstein:2014yza}. It thus seems that, despite being non-covariant, the figure or merit serves as a reliable estimate for a qualitative measure of near-horizon (non-)adiabaticity. In agreement with the first quantized method for computing string production, non-adiabicity occurs for large energy per unit mass of the D-branes. A curious feature, however, is that in this limit the relative boost between the pair of D-branes - which served as the original motivation for studying possible non-adiabatic effects - drops out!

So far we discussed evaluating the figure of merit at the horizon as this kind of non-adiabaticity would most closely resemble a firewall in the present context. Non-adiabatic string production effects are, however, not restricted to the near-horizon region and, indeed, they grow towards the singularity. Because Gullstrand-Painlev\'{e} coordinates include the singularity, the locally estimated figure or merit ceases to be a valid measure for non-adiabaticity at late times. On nice slices, because they are smooth everywhere and avoid the singularity by construction, we can reliably use the figure of merit to estimate non-adiabaticity so long as it remains small enough for the WKB approximation to be applicable. 

Contrary to the original expectation we, however, find non-adiabaticity in the nice slicing at very late time where the time-dependent mass (or frequency) goes to zero leading to a breakdown of the WKB approximation. The reason for the breakdown of the WKB approximation turns out to be due to a not so nice feature of nice slices: accumulation. Because of the existence of a last nice slice (recall that nice slices are constructed to avoid the curvature singularity) there has to be an accumulation of slices somewhere in the black hole geometry. While this accumulation effect cannot affect the local physics it will lead to a non-adiabatic effect in the global pair production rate between the asymptotic {\it in} and {\it out} states. To illustrate this effect consider, as an analogy, a putative transition in a harmonic oscillator between an excited state with definite non-zero energy to a state with zero energy. The resulting divergence in the average number of excited states is analogous to the non-adiabatic 
string production we find in a nice slicing of black holes.\footnote{We thank Gary Horowitz and Mark Srednicki for discussion on this issue.} We can explain this breakdown rather explicitly due to a nice feature of our nice slice construction which is that we can model the time-dependent frequency by an analytic function that takes the form of a hyperbolic tangent.\footnote{This functional form is very powerful: when the frequency is an hyperbolic tangent the particle creation problem (strings in our case) can be solved exactly, {\it i.e.} without the need to resort to the WKB approximation (\emph{e.g.}, see~\cite{Birrell:1982ix}). We will make use of this property in~\S~\ref{ssec:Schwarzschildscattering} when we attempt to solve the string production problem in the Schwarzschild geometry directly using a real-time method.}

Finally one may worry that a large relative boost between the D-branes may produce strings with size larger than the horizon scale which would invalidate an analysis that makes use of the Hagedorn density of states computed in flat space. To address this issue we compute the size of the open strings stretched between the D-branes. Our findings suggest that there exists a range of parameters where the size of the produced strings remains small despite having large boosts. This justifies our non-adiabaticity analysis.

\bigskip

The organization of this paper is as follows. In~\S\ref{sec:stringproduction} we review some of the methods for computing the production rate of open strings: the (non-covariant) real-time estimate $\omega^2/\dot{\omega}$, the first quantized formalism introduced in~\cite{Silverstein:2014yza,Joeunpublishednote} and the real-time string production method using asymptotic {\it in} and {\it out} states. In~\S\ref{sec:nicetrajectory} we introduce the notion of nice slices, first for the black hole geometry in~\S~\ref{ssec:Schwarzschildslice} and then for the near-horizon region (Rindler space) in~\S~\ref{ssec:flatslice}, and compute the relevant D-brane trajectories in~\S~\ref{ssec:Schwarzschildtrajectory} and~\S~\ref{ssec:flattrajectory}. In~\S\ref{sec:flatproduction} and~\S\ref{sec:Schwarzschildproduction} we compute the rate of open string pair production in, respectively, Rindler space and the entire Schwarzschild geometry. In~\S\ref{sec:size} we compute the size of the produced 
strings. 
Finally in~\S\ref{sec:discussion} we conclude with a discussion of our results, their relation to other works in the literature and some open questions.

\section{Open strings pair production and near-horizon boost}\label{sec:stringproduction} 

In this section we review some basic facts about strings in curved backgrounds and the methods for computing the rate of open string pair production. We then discuss the dynamical thought experiment first considered in~\cite{Silverstein:2014yza} that connects possible non-adiabatic string-theoretic effects to the physics of infalling observers in a Schwarzschild black hole.

\subsection{String production in curved backgrounds}

We start from the Polyakov action describing the motion of a string in a curved background 
\begin{equation}
S= -\frac{1}{\alpha'} \int d\tau d\sigma \sqrt{-h} h^{\alpha \beta} G_{MN}\partial_\alpha X^M \partial_\beta X^N\,.
\end{equation}
The spacetime equation of motion for the string wave function $|\Psi\rangle$, related to the Hamiltonian constraint $T_{\alpha\beta}=  -\frac{1}{\sqrt{-h}}\frac{\delta S}{\delta h^{\alpha \beta}}=0$ that enforces reparametrization invariance on the worldsheet, is  
\begin{equation}
\label{eq:waveeq}
T_{\alpha\beta} |\Psi\rangle =0\,.
\end{equation}
For an arbitrary background metric this yields, in general, a complicated equation. However, for spacetimes exhibiting enough symmetry and for simple enough string profiles, this wave equation is simple enough to be solved using WKB techniques.

Take for example an open string in Milne space
\begin{equation}
ds^2=-dT^2+T^2 dY^2 +dX_\perp^2\,,
\end{equation}
and consider the Ansatz for the string profile
\be
T(\tau, \sigma) = T(\tau), \quad Y(\sigma,\tau) = \frac{\eta \sigma}{\pi},\quad X_\perp(\tau,\sigma)= \frac{b_\perp \sigma}{\pi}.
\ee
In this case, the wave equation~\eqref{eq:waveeq} yields
\begin{equation}
\label{eq:ho}
 \ddot{\Psi}(T) + \omega^2(T)\Psi(T)=0\,,
\end{equation}
where $\dot{}\equiv d/d T$ and $\omega(T)=\frac{1}{\alpha'}\sqrt{\eta^2 T^2 + b_\perp^2}$. This describes a simple harmonic oscillator with a time-dependent frequency $\omega(T)$.

In this article, all of our equations describing strings in different background metrics will take the form~\eqref{eq:ho}, albeit with different time-dependent frequencies $\omega$.\footnote{See \cite{Lawrence:1995ct} for a similar approach based on string field theory computations.} Also, if the Ansatz for the string profile stretches the string along the symmetric directions of the spacetime metric, there is no loss of generality by assuming such an Ansatz. It is worth noting that, after imposing the Hamiltonian constraint which converts the Polyakov action into the Nambu-Goto action, and integrating over the $\sigma$ coordinate for our string profile, the action will always take the form:
\begin{equation}
S =-\frac{1}{\alpha'} \int d\tau d\sigma \sqrt{-{\rm det} G_{MN} \partial_\alpha X^M \partial_\beta X^N}  \equiv \int \omega(T)\, dT \,.\label{eq:action}
\end{equation}
In the time regimes where $\dot \omega/\omega^2 \ll 1$, the wave equation can be solved, to good accuracy, by the WKB wave function
\begin{equation}
\label{eq:WKBfunction}
\Psi(T) \sim \frac{1}{\sqrt{2\omega(T)}}e^{-i \int^T \omega(T') dT'}.
\end{equation}
Note that the phase in this solution is precisely the action in \eqref{eq:action}.

Consider now systems in which $\dot \omega/\omega^2$ vanishes for $T \to \pm \infty$. If in the infinite past we start with the pure positive-frequency solution $\Psi_{\rm in}(T)$ given by \eqref{eq:WKBfunction}, in the far future the system will also be described by a WKB function, but now with a mixture of positive and negative frequencies, \emph{i.e.},
\begin{equation}
\label{eq:posnegfreq}
\Psi_{\rm out}(T) \sim \frac{\alpha}{\sqrt{2\omega(T)}}e^{-i \int^T \omega(T') dT'} + \frac{\beta}{\sqrt{2\omega(T)}}e^{i \int^T \omega(T') dT'}.
\end{equation}
If the system starts in the vacuum state $|0_{\rm in}\rangle$, the number of strings created after the system has completely evolved is given by the Bogoliubov coefficient $|\beta|^2=\langle 0_{\rm in}| a_{\rm out}^\dagger a_{\rm out}|0_{\rm in}\rangle$. For each oscillator level $n$, strings states have a degeneracy  $\rho(n)$ given by the number of partitions of $n$, thus, the total number of produced strings is
\begin{equation}\label{eq:Ntotgeneral}
 N_{tot} =\sum_{n=1}^{\infty} \rho(n) |\beta_n|^2\,.
\end{equation}
There are various ways to compute the associated Bogoliubov coefficient $\beta$ in the WKB approximation. We now summarize the methods that we will later use for computing the rate of open string production. For more details we refer to~\cite{Silverstein:2014yza}.

\paragraph{Real-time method.}
The real-time method for computing $|\beta|^2$ corresponds to solving the wave equation~\eqref{eq:ho} \emph{exactly}, and then matching such exact solution with the approximate ones given in \eqref{eq:WKBfunction} and \eqref{eq:posnegfreq} in the far past and future respectively. Gluing these solutions gives the Bogoliubov coefficients $\alpha$ and $\beta$.
In particular $|\beta|^2$ can be extracted by continuing the result $T\to e^{-i\pi} T$:\footnote{This trick also works for the approximate WKB solution~\eqref{eq:WKBfunction}.}
\begin{equation}
 \Psi(e^{-i\pi} T) = |\beta|^2 \Psi(T)\,.
\end{equation}
In general the wave equation has to be solved numerically. 

\paragraph{Saddle-point method.}
 The first quantized method of \cite{Silverstein:2014yza,Joeunpublishednote} to calculate string production uses a saddle-point approximation as in \cite{Chung:1998bt,Gubser:2003vk}. The Bogoliubov coefficient can be computed by a suitable contour integral of $\omega (T)$ in the complex $T$ plane. Namely it is given by   
 \begin{equation}
\label{eq:beta_tunnel}
  |\beta|^2 \sim e^{-2 {\rm Im} S}\,,
 \end{equation}
with the action
\begin{equation}
 S=\int_{T_i}^{T_f} \omega(T) dT \equiv \int_{T_i}^{T_*} \omega(T) dT +\int_{T_*}^{T_f} \omega(T) dT\,.
\end{equation}
where $T_{i} $ and $T_{f}$ label the initial and final time slices.
The string production saddle point of the worldvolume path integral (with path integration over $dT_*$) is captured by a contour that winds around the branch point $T_*$ where the frequency vanishes $\omega(T_*)=0$ thus implementing the Hamiltonian constraint. The exponentially decaying form of \eqref{eq:beta_tunnel} is of the tunnelling amplitude-type we alluded to in the introduction section.
The imaginary part of the action can be obtained by integrating vertically from the nearest point on the real axis to the branch cut
\begin{equation}
\label{eq:ImS}
 {\rm Im} S \approx -2i \int_{T_{re}}^{T_*} \omega(T) dT\,,
\end{equation}
where the nearest singularity is at $T_*=T_{re}+i T_{im}$. 
The length of the contour is of order $\omega/\frac{d\omega}{dT}$ and so the imaginary part is of order $\omega^2/\frac{d\omega}{dT}$. For more details see~\cite{Silverstein:2014yza}.

\paragraph{Real-time estimate.}
We can also estimate $|\beta|^2$ with the figure of merit
\begin{equation}\label{eq:betaestimate}
 |\beta|^2 \sim e^{-\left|\omega^2/\dot{\omega}\right|}\,.
\end{equation}
which is valid only when $\dot{\omega}/\omega^2 \ll 1$, \emph{i.e.}, when we are in the adiabatic regime. 

Equation~\eqref{eq:betaestimate}, more concretely the dependence of $\omega$ on $T$, appears to be at odds with the definition of the Bogoliubov coefficient $\beta$ being a constant. However, the real meaning of equation~\eqref{eq:betaestimate} is that it relates the transition between two adiabatic vacua, where the \emph{in} vacuum is defined in the far past ($T \to -\infty$) and the \emph{out} vacuum is defined at an arbitrary intermediate \emph{adiabatic} vacuum at time $T$.\footnote{It would be interesting to apply to our problem the universal definition of \emph{time-dependent} particle number recently proposed by~\cite{Dabrowski:2016tsx}.}

When it is not possible to define an asymptotic \emph{out} region because $\dot{\omega}/\omega^2$ not only fails to be small at future infinity, but even blows up, we will rely on the estimate \eqref{eq:betaestimate} to infer the number of strings produced in the regions where $\dot{\omega}/\omega^2$ remains small.

\subsection{The Gedankenexperiment}

We now explain the Gedankenexperiment, first introduced in~\cite{Silverstein:2014yza}, in which open strings are produced between D-branes moving in a black hole background. For definiteness, we consider a four dimensional  Schwarzschild black hole, whose metric is given by 
\begin{equation}
ds^2=-f(r) dt^2 + \frac{dr^2}{f(r)} + r^2 d\Omega^2\,, \qquad f(r)=1-\frac{r_0}{r}\,.
\label{eq:Schwarzschild}
\end{equation}
In this background we prepare a pair of D0 branes at fixed radial coordinate $r=R$. First, at $t=0$ we drop the first of the D0 branes which follows the trajectory $(t(r),r)$ of a particle freely falling into the black hole. As the first D0 brane is falling, the open string stretching between two D0 branes is expanding. Next, at $t=\Delta t$ we release the second D0 brane which follows the time-shifted trajectory $(t(r) +\Delta t,r)$. 
Since the time shift $t \rightarrow t+ \Delta t $ in Schwarzschild coordinates $(t,r)$ corresponds to the boost $X^{\pm} \rightarrow e^{\pm \Delta t}X^{\pm}$ in the Kruskal coordinates $(X^+,X^-)$, from the point of view of the first D0 brane, the second D0 brane is highly boosted in the large $\Delta t$ limit.  

Based on this simple observation and the result of~\cite{Bachas:1995kx} that the relative boost of two D-branes leads to the enhanced production of open strings stretching between them, one may expect non-adiabatic open string pair production that cannot be captured by effective field theory on the fixed black hole geometry.
Rephrasing the Gedankenexperiment in a ``nice'' slicing of the four dimensional Schwarzschild black hole we will investigate the possibility of such a boost-enhanced non-adiabatic effect. In doing so we will uncover a rather ugly feature of nice slices which presents an obstruction to the use of effective field theory in describing the entire black hole evolution process.

\section{Nice slices and D-brane trajectories}\label{sec:nicetrajectory}

\begin{figure}[ht!]
\centering{
\subfigure[Nice slices in Schwarzschild black holes.]{
\includegraphics[width=.42\textwidth]{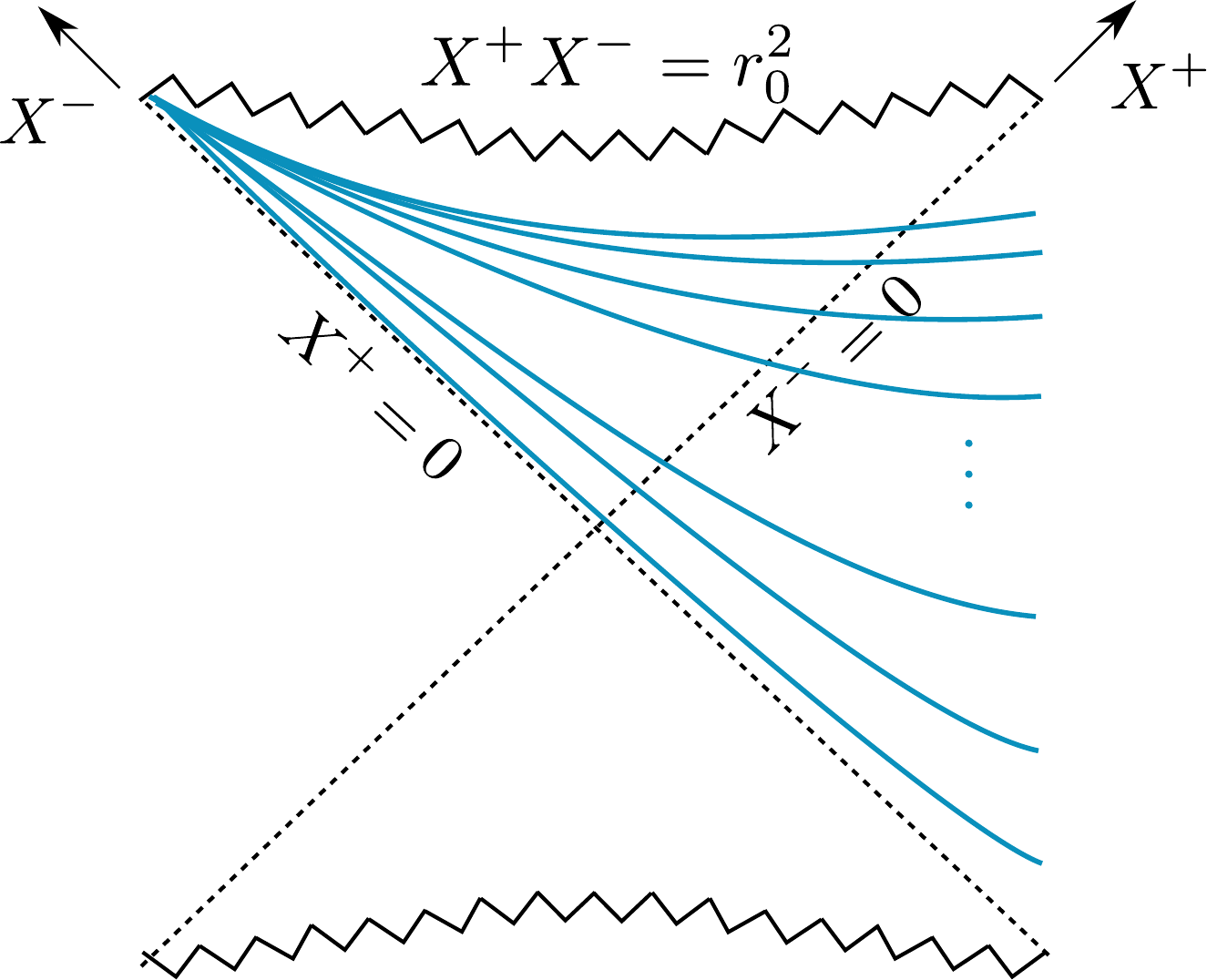}
\hspace{1cm}}       
\subfigure[Nice slices in flat spacetime.]{
\includegraphics[width=.42\textwidth]{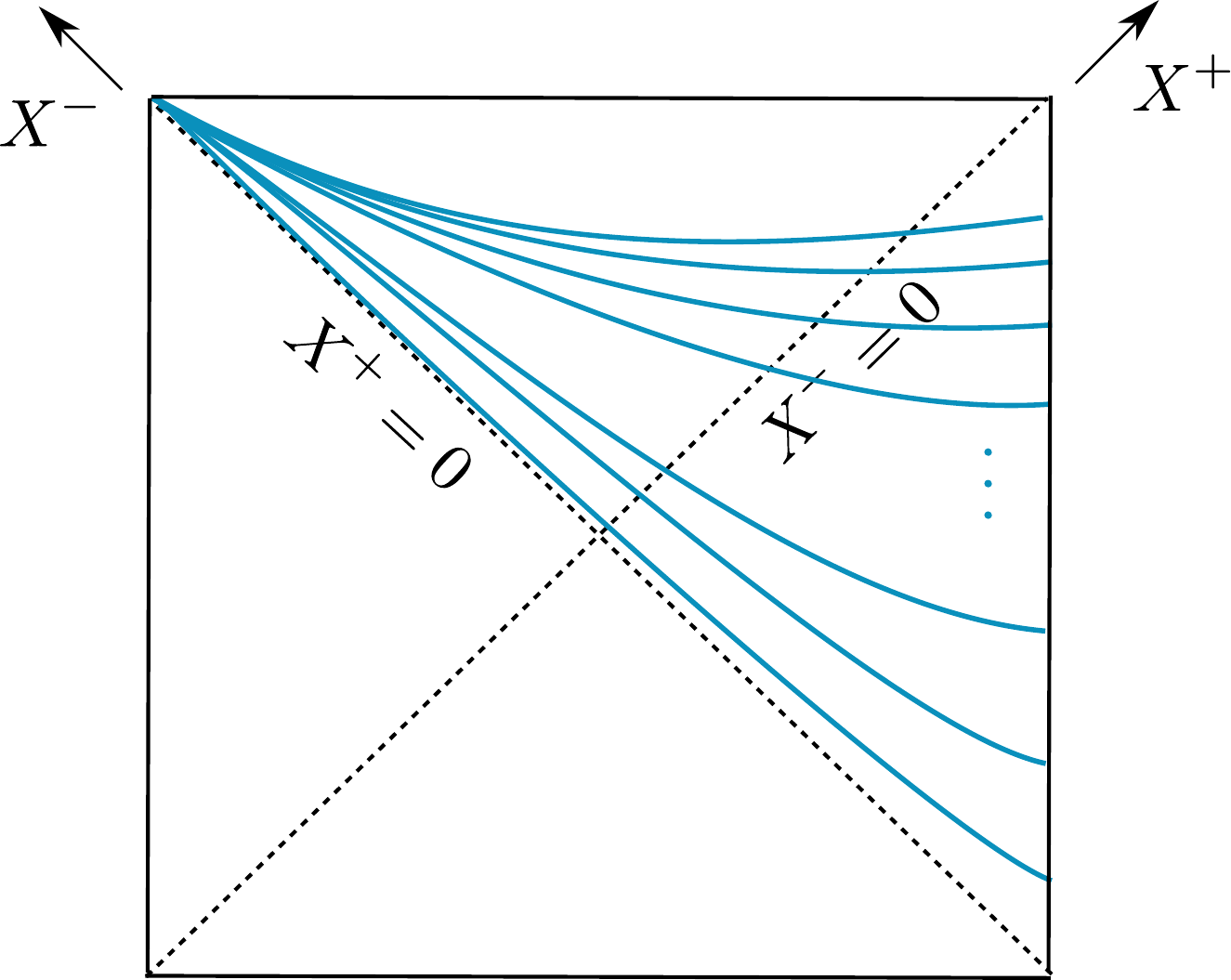}}
}
\caption{\small Depicted in blue is a nice slicing of (a) Schwarzschild black hole and (b) flat spacetime.
\label{fig:niceslices}}
\end{figure}

In this section we will introduce the notion of ``nice slices''. These foliate spacetime by a family of surfaces that smoothly enter the horizon, avoid any singularities and are everywhere well behaved (they are smooth, there are no large curvatures or accelerations, any matter must be moving with modest velocity in the local frame defined by the nice slice) so that the physics on the slice has a well-defined semi-classical description~\cite{Lowe:1995ac,Polchinski:1995ta}. We first introduce these slices for Schwarzschild black holes following~\cite{Giddings:2012bm} and then repeat the same construction for Rindler spacetime describing the near-horizon region. See Figure~\ref{fig:niceslices} for an illustration. We also discuss the trajectories of free falling D-branes in the geometries which we will use in the calculations of open string pair production in~\S~\ref{sec:flatproduction} and~\S~\ref{sec:Schwarzschildproduction}.

\subsection{Nice slicing of Schwarzschild black holes}\label{ssec:Schwarzschildslice}

Let us  illustrate the construction of nice slice in a four dimensional Schwarzschild black hole with metric~\eqref{eq:Schwarzschild} following~\cite{Giddings:2012bm}. First we switch from Schwarzschild coordinates $(t,r)$ to tortoise coordinates $(t,r_*)$:
\begin{equation}
 ds^2 = f(r) (-dt^2+dr_*^2)+r^2(r_*) d\Omega^2\,,\qquad \frac{dr_*}{dr}=\frac{1}{f(r)}\,,
\end{equation}
where $r_*=r+r_0 {\rm log}\left|1-\frac{r}{r_0}\right|$
in which the horizon $r=r_0$ is at $r_*=-\infty$. We then go to Kruskal coordinates
\begin{equation}
 X^{\pm} = \pm r_0 e^{\pm x^{\pm}/2 r_0}\,, \qquad x^\pm = t\pm r_*\,.
\end{equation}
The singularity is at $X^+ X^-=r_0^2$ and the future horizon is at $X^-=0$. We use spatial slicings for a given time $T$ defined by
\begin{equation}
 X^+ (X^- + e^{-T/r_0} X^+) = R_c^2\,, \qquad \text{for fixed}\; R_c>0\,.\label{eq:slicing}
\end{equation}
Different choices of $R_c$ produce the following sets of slices:
\begin{itemize}
 \item $R_c=0$: Schwarzschild slicing (slices never cross the horizon)
 \item $0<R_c<r_0$: nice slicing (slices smoothly enter the horizon but avoid the singularity)
 \item $R_c>r_0$: natural slicing (slices smoothly enter the horizon but hit the singularity)
\end{itemize}
 The inside of the black hole is described in nice slice time until $T=\infty$ which is at the (finite) radius
\begin{equation}
 r_\infty= r_0 \left(1+{\rm Productlog} \left[-\frac{1}{e} \left(\frac{R_c}{r_0}\right)^2\right]\right)\,.\label{eq:lastradiusniceslice}
\end{equation}

In terms of the Schwarzschild coordinates $(t,r)$, the nice slice time $T$ is obtained from~\eqref{eq:slicing}, yielding
\begin{equation}\label{eq:Tofr}
 T=t+r_*-r_0 {\rm log}\left[\left(\frac{R_c}{r_0}\right)^2- e^{r/r_0}(1-r/r_0)\right]\,.
\end{equation}
Note that this expression is valid for both outside and inside the horizon. 

\subsection{Trajectories in Schwarzschild black holes}\label{ssec:Schwarzschildtrajectory}
Here we summarize trajectories of free particles, or D0 branes, in the Schwarzschild black holes. 
The conserved energy of the trajectory is given by 
\begin{equation}
 E=mf(r) \dot{t} = m \frac{\sqrt{f(r)}}{\sqrt{1-\frac{\left(\frac{dr}{dt}\right)^2}{f(r)^2}}}\,,
\end{equation}
where $E$ and $m$ are the energy and mass of the particle. The trajectory $t(r)$ is obtained by integrating
\begin{equation}
 \frac{dr}{dt} = -f(r) \sqrt{1-\frac{f(r)}{C^2}}\,,\quad \text{where} \quad C\equiv E/m\,. \label{eq:drdtSchwarzschild}
\end{equation}

We consider a symmetric setup where the first D0 brane moves along $(t(r),r)$ while the second D0 brane moves along $(t(r) +\Delta t,r)$. Because of the time translation symmetry of the spacetime, if $(t(r),r)$ is a particle trajectory, $(t(r) +\Delta t,r)$ also satisfies (\ref{eq:drdtSchwarzschild}). Note that the log term in $t(r)$, which makes it blow up at the horizon, gets precisely canceled by the term $r_*$ in~\eqref{eq:Tofr}, thus, making $T(r)$ a smooth parametrization for physics across the horizon.

\subsection{Nice slicing of Rindler space}\label{ssec:flatslice}

Recall that the near-horizon region of a Schwarzschild black hole in the limit $\rho^2=4 r_0 (r-r_0) \approx 0$ is described by flat spacetime in Rindler coordinates
\begin{equation}
 ds^2=-\frac{\rho^2}{4r_0^2} dt^2+d\rho^2+dX_\perp^2\,.
\end{equation}
To study string production between pairs of boosted D-branes crossing the Rindler horizon in flat spacetime we adapt the construction from the previous section. First we change to light-cone coordinates
\begin{equation}
 X^\pm = \pm \rho e^{\pm t/2r_0}\,.
\end{equation}
Nice slices are then defined by
\begin{equation}
 X^+(X^-+e^{-T/r_0} X^+)=R_c^2\,.
\end{equation}
The nice slice time in flat space is given by\footnote{We denote the nice slice time in the Schwarzschild and Rindler geometry by the same letter, $T$, but this should not result in any confusion as the discussions will be well separated.}
\begin{equation}
 T= t-r_0 {\rm log} \left[\left(\frac{R_c}{\rho}\right)^2+1\right]\,.
\end{equation}

\subsection{Trajectories in Rindler space}\label{ssec:flattrajectory}

We will now specify the Rindler trajectory $t(\rho)$. The conserved energy of the trajectory is given by
\begin{equation}
 E=m \frac{\rho^2}{4r_0^2} \dot{t}=m \frac{\frac{\rho}{2r_0}}{\sqrt{1-\left(\frac{2r_0}{\rho}\right)^2\left(\frac{d\rho}{dt}\right)^2}}\,.
\end{equation}
The trajectory $t(\rho)$ is obtained by integrating
\begin{equation}
\frac{dt}{d\rho}=-\frac{2r_0}{\rho} \frac{1}{\sqrt{1-\frac{\rho^2}{4r_0^2}\frac{1}{C^2}}}\,, \quad \text{where}\quad C\equiv E/m\,.
\label{eq:Rindlertrajectory2}
\end{equation}
Note that $C \ge1$ and the range of $\rho$ is restricted to $\rho\leq R=2r_0 C$.
The trajectory of the first D0 brane is given by
\begin{equation}
 t(\rho)=2 r_0 \left(-\log \left[\rho \right]+ \log \left[1+\sqrt{1-\frac{\rho^2}{4 r_0^2}\frac{1}{C^2}}\right]\right)\,.
\end{equation}
The trajectory of the second D0 brane is obtained by $t(\rho) + \Delta t$. 
This completely specifies the radial trajectory in nice slice time $T(\rho)$.

\section{String production on nice slices in Rindler space}\label{sec:flatproduction}

We first illustrate the methods for computing open string pair production discussed in~\S~\ref{sec:stringproduction} for boosted branes in a nice slicing of flat spacetime in Rindler coordinates matching to the results of~\cite{Silverstein:2014yza,Bachas:1995kx}. 
 
We want to compute the production of open strings between a pair of D0 branes moving on constant velocity trajectories $u=u_0+ V v$ in Minkowski space 
\begin{equation}
 ds^2=-dv^2+du^2+dX_\perp^2\,.
\end{equation}
The Minkowski and Rindler coordinates are related by
\begin{equation}
 v=\rho \sinh\left(\frac{t}{2r_0}\right)\,, \quad u=\rho \cosh\left(\frac{t}{2r_0}\right)\,.
\end{equation}
We consider a pair of D0 branes, with the open string endpoints at $\sigma=0,\pi$, embedded in Rindler space as\footnote{We will consider a different string profile in~\S~\ref{ssec:yetanother} below.}
\begin{equation}
 X^M=(t,\rho,X_\perp)\,, \quad \partial_\tau X^M=(\dot{t},\dot{\rho},0)\,, \quad \partial_\sigma X^M=(\frac{\Delta t}{\pi},0,\frac{b_\perp}{\pi})\,,
 \end{equation} 
 where $\Delta t$ and $b_\perp$ denote the separation of the two D0 branes in, respectively, time and transverse space.
The string action~\eqref{eq:action} becomes
\begin{equation}
 S=-\frac{1}{\alpha'} \int d\rho \sqrt{-b_\perp^2\left(1-\frac{\rho^2}{4r_0^2} \left(\frac{dt}{d\rho}\right)^2\right) + \Delta t^2 \frac{\rho^2}{4r_0^2}}\,.
\end{equation}
Using
\begin{equation}
 \rho=\sqrt{u^2-v^2}\,, \quad \frac{t}{2r_0} =\frac{\Delta t}{2r_0} \frac{\sigma}{\pi} + {\rm log}\sqrt{\frac{u+v}{u-v}}
 \end{equation}
we get for the D0 brane trajectory in Rindler space
\begin{equation}
\frac{dt}{d\rho}=-\frac{2r_0}{\rho} \frac{u_0}{\sqrt{u_0^2+(V^2-1)\rho^2}}\,.\label{eq:Rindlertrajectory1}
\end{equation}
Comparing this with~\eqref{eq:Rindlertrajectory2}, and using $C^2=1/(1-V^2)$ and $u_0=2r_0$, the string action becomes
\begin{equation}
 S=-\frac{1}{\alpha'} \int d\rho \frac{\rho}{2r_0} \sqrt{\frac{b_\perp^2}{C^2-\frac{\rho^2}{4r_0^2}} + \Delta t^2}\equiv \int d\rho \omega(\rho)\,.  
\end{equation}
The string oscillator level $n$ is captured by substituting $b_\perp^2 \to b_\perp^2 +\alpha' n $.
In a nice slicing of flat spacetime with respect to the nice slice time coordinate $T$ defined in~\S~\ref{ssec:flatslice} the string action at oscillator level $n$ becomes
\begin{equation}
S_n =\int dT \omega_n(T)\,, \quad \text{with}\quad  \omega_n(T) = -\frac{1}{\alpha'}  \frac{\rho}{2r_0} \sqrt{\frac{b_\perp^2+ n \alpha'}{C^2-\frac{\rho^2}{4r_0^2}}+\Delta t^2} \frac{d\rho}{dT}\,.\label{eq:niceactionflat}
\end{equation}
We can now compute the Bogoliubov coefficient for string production on nice slices in flat spacetime using the methods of~\S~\ref{sec:stringproduction}.

\subsection{Real-time estimate}\label{ssec:flatestimate}
We can use the figure of merit (\ref{eq:betaestimate}) to estimate the level of non-adiabaticity at the Rindler horizon. For large oscillator levels $n$ the density of states behaves as $\rho(n) \sim e^{\sqrt{8\pi^2 n}}$ and the total number of produced strings~\eqref{eq:Ntotgeneral} becomes
 \begin{equation}\label{eq:Ntotestimatelargen}
 N_{tot} \approx \sum_{n \gg 1}^{\infty} e^{\sqrt{8\pi^2 n} -|\omega_n^2/\dot{\omega}_n|} \,,
 \end{equation}
with $\omega_n$ given by~\eqref{eq:niceactionflat}. Using the expression for $T(\rho)$ from~\S~\ref{ssec:flatslice} we get
\begin{equation}
 \frac{d\rho}{dT}=\frac{\frac{\rho}{2r_0}}{\frac{R_c^2}{R_c^2+\rho^2}-\frac{1}{\sqrt{1-\frac{\rho^2}{4r_0^2 }\frac{1}{C^2}}}}\,.
\end{equation}
Note that while $d\rho/dT$ diverges at $\rho=0$, the frequency $\omega(\rho)$ vanishes at the Rindler horizon so that $\omega(T)=\omega(\rho(T)) d\rho/dT$ remains finite.
The figure of merit evaluated at the Rindler horizon $\rho=0$ is
\begin{equation}
  \left|\frac{\omega_n^2}{\dot{\omega}_n}\right|_{horizon}\hspace{-0.5cm}=\frac{4 C r_0 R_c^2 \left(8 C^2 r_0^2+R_c^2\right) \left(b^2+\alpha'  n +C^2 \Delta t^2\right)^{3/2}/\alpha' }{(b^2+\alpha'  n) \left(1 6C^2 r_0^2( 8C^2 r_0^2+  R_c^2)-R_c^4\right)+C^2 \Delta t^2 \left(128 C^4 r_0^4-3 R_c^4\right)}\,.
\end{equation}
In the large $C$ limit the contribution from the high $n$ oscillator levels becomes
\begin{equation}
\label{eq:Rindlerestimate}
 \left|\frac{\omega_n^2}{\dot{\omega}_n}\right|_{horizon}\sim\frac{\sqrt{n}}{\sqrt{\alpha'} C}\frac{R_c^2}{4 r_0}\,.
 \end{equation}
The total number of open strings $N_{tot}$ produced at the Rindler horizon is non-adiabatically enhanced for
\begin{equation}
\label{eq:boundCestimate}
  C\gtrsim\frac{R_c^2}{8 \sqrt{2}\pi r_0 \sqrt{\alpha'}}\,.
\end{equation}
Recall that the estimate $|\beta|^2\sim e^{-|\omega^2/\dot \omega|}$ is only valid for $|\omega^2/\dot \omega| \gg 1$, which is indeed satisfied for $C$ of the order given in \eqref{eq:boundCestimate} as long as $n$ is large.

To get a better understanding of this non-adiabaticity regime recall that in flat space we can write $C\equiv E/m=1/\sqrt{1-V^2}$. So, a large energy per unit mass, $C\gtrsim \mathcal{O}(r_0/\sqrt{\alpha'})$, corresponds to a large D-brane velocity. It may not be too surprising that string production is non-adiabatically enhanced in the extreme limit where the D-branes move with a velocity close to the speed of light, $V\to1$,\footnote{Already in~\cite{Bachas:1995kx} it was noted that string-production is non-adiabatic in that limit; the speed of light was related to the critical value of the electric field in the T-dual description of open-string pair production.} in particular, since in ultrarelativistic limit $C\gg1$ the geometry is described by a gravitational shockwave.\footnote{This observation is also relevant in the black hole case discussed in section~\S~\ref{ssec:Schwarzschildsaddle}. We would like to thank Joe Polchinski and Samir Mathur for discussion on this point.} The estimate also suggests 
non-adiabaticity for more moderate values of $C$, however, these values are coordinate and slice dependent.

To establish whether the non-adiabatic enhancement of string production is confined to the Rindler horizon we can evaluate the estimate farther outside, say at $\rho=\gamma r_0$ for some $0<\gamma<2C$. At large oscillator level $n$, the large $C$ limit gives:
\begin{equation}
 \left|\frac{\omega_n^2}{\dot{\omega}_n}\right|_{outside}\sim\frac{r_0\sqrt{n}}{4\sqrt{\alpha'} C}\left(\frac{R_c^2}{ r_0^2}+\gamma^2 \right)\,.
 \end{equation}
Hence, for $C\gtrsim \frac{r_0\left(\frac{R_c^2}{r_0^2}+\gamma^2\right)}{2\sqrt{2}\pi\sqrt{\alpha'}}$ string production is again enhanced. The non-adiabaticity is thus not confined to the region near the Rindler horizon, at least in the large $C$ limit.

\subsection{Saddle point method}\label{ssec:flatdsaddle}
To capture string production in the first quantized method using a saddle point approximation we need to choose the relevant contour and integrate around the branch points of $\omega(T)$. Since $\frac{d\rho}{dT}$ is smooth in the range $0\leq \rho\leq R=2r_0 C$ we can replace the $T$ integral by the $\rho$ integral:
\begin{equation}
 S=\int_{T_i}^{T_f} \omega(T) dT = \int_{\rho_i}^{\rho_f} \omega(\rho) d\rho\,.
\end{equation}
The integrand has a branch cut between an integrable square root singularity at $\rho_{*,1}=\pm 2 r_0 C$ and a branch point at
\begin{equation}\label{eq:flatbranchpoint}
  \rho_{*,2}=\pm \frac{2r_0}{\Delta t}\sqrt{b_\perp^2 + C^2 \Delta t^2}\,,
\end{equation}
 where $\omega(\rho)$ vanishes. 
In the nice slice coordinate $T$, we also have a zero of $\omega(T)$ at $T=\infty$ in addition to~\eqref{eq:flatbranchpoint} but in this region nice slices pile up and the WKB approximation breaks down.\footnote{We will discuss this effect in detail in~\S~\ref{ssec:Schwarzschildscattering}.}
The contribution to the imaginary part of the action capturing string production comes from the part of the contour that goes around the cut between the branch point $\rho_{*,1}$ and the singularity $\rho_{*,2}$. The result is
\begin{equation}\label{eq:flatImAction}
 S=-\frac{1}{\alpha'}\int d\rho \frac{\rho}{2r_0} \sqrt{\frac{b_\perp^2}{C^2-\frac{\rho^2}{4r_0^2}} + \Delta t^2}=i\frac{\pi b_\perp^2}{2 \alpha' \eta}\,,
\end{equation}
where we defined the boost $\eta\equiv \Delta t/2 r_0$. The Bogoliubov coefficient is
\begin{equation}\label{eq:flatBogoliubov}
 |\beta^2|\sim e^{-\frac{\pi b_\perp^2}{\alpha' \eta}}\,.
\end{equation}
Hence, there is a non-adiabatically \emph{boost} enhanced production of open strings between D0 branes moving on relatively relativistically boosted trajectories in a nice slicing of flat spacetime. The result~\eqref{eq:flatBogoliubov} agrees with the one found in~\cite{Silverstein:2014yza}.\footnote{In~\cite{Silverstein:2014yza} this calculation was performed in Milne space $ds^2=-dT^2+T^2 dY^2+dX_\perp^2$ which is related to Rindler space by $T=i \rho$ and $e^{2Y}=-e^{t/r_0}$.} This is as expected since the result should not depend on the slicing we choose.

\subsection{Another string profile}\label{ssec:yetanother}

To see whether non-adiabatic string production is a universal phenomenon in the boosted system or not, we now consider a different string profile and calculate the production rate in the nice slice coordinates $(T,X^{+},X^{\perp})$. In these coordinates the flat metric is given by 
\begin{align}
ds^2&=-dX^{+}dX^{-} +dX_{\perp}^2 \\
&=-\frac{e^{-\frac{T}{r_{0}}}X^{+}}{r_{0}}dT dX^{+}+ \left(e^{-\frac{T}{r_{0}}} +\frac{R_{c}^2}{(X^{+})^2} \right) (dX^{+})^2+dX_{\perp}^2 .
\end{align}

The string profile we consider is 
\begin{equation}\label{eq:niceprofile}
T(\tau,\sigma) =T(\tau) +\frac{\Delta T}{\pi} \sigma, \quad X^{+} (\tau,\sigma) = \frac{a}{\pi} \sigma, \quad X^{\perp}(\tau,\sigma) =\frac{b_{\perp}}{\pi} \sigma,
\end{equation}
where $T(\tau)$ denotes the trajectory of the first D0 brane and $a$ denotes the separation along the null coordinate $X^{+}$. 
Note that this profile is different from the one considered in the previous section, and is perhaps a more natural profile to consider with respect to the nice slices.  

The Nambu-Goto action for the string profile~\eqref{eq:niceprofile} is given by 
\begin{align}
S&=-\frac{1}{\alpha '} \int d\tau d \sigma  \;e^{-\frac{T(\tau)}{r_{0}}} \left(\frac{a X^{+}(\tau,\sigma) }{2r_{0}\pi} \right) \frac{d T}{d \tau},  \nonumber\\
&=-\frac{a b_{\perp}}{2\alpha' r_{0}} \int  e^{-\frac{T}{r_{0}}} \;  dT\,.\label{eq:niceprofilesaddle}
\end{align}
Note that the action does not depend on the energy of the D0 brane  $C\equiv E/m$ or the boost $\Delta T$. 
The structure of the integrand $\omega(T)$ is so simple that we immediately see that there is no branch cut or singularity. Naively, we would therefore conclude that there is no string production. 

On the other hand, if we compute the estimate (including the oscillator level $n$) we get:
\begin{equation}
\frac{\omega_{n}^2}{\dot{\omega}_{n}} =-\frac{a\sqrt{b_{\perp}^2+\alpha' n}}{2\alpha'} \; e^{\frac{T}{r_{0}}}.\label{eq:estimate}
\end{equation}
The adiabaticity figure $\dot \omega / \omega^2$ diverges for late times $T\gg r_0$ and the WKB approximation breaks down. Hence, in the saddle point method we should bound the range of integration in~\eqref{eq:niceprofilesaddle} by $T_c$ beyond which it is no longer justified to use the WKB approximation. Then, the absence of an imaginary part of the action is consistent with the adiabatic estimate~\eqref{eq:estimate}. 

At large oscillator level $n$ the estimate becomes
\begin{equation}
\frac{\omega_{n}^2}{\dot{\omega}_{n}} \sim e^{\frac{T}{r_{0}}} \frac{a \sqrt{n}}{\sqrt{\alpha'}}.\label{eq:estimateniceprofile}
\end{equation}
At fixed $T$ for
\begin{equation}
\frac{a}{\sqrt{\alpha'}} \sim \mathcal{O}(1)
\end{equation}
there is again non-adiabatically enhanced string production.

\section{String production on nice slices in Schwarzschild}\label{sec:Schwarzschildproduction}
In this section we discuss the main problem of interest, namely string production in the background of a Schwarzschild black hole. 
As in flat spacetime we focus on open string pair production from two boosted D0 branes that follow symmetric trajectories in the Schwarzschild geometry~\eqref{eq:Schwarzschild} with embedding
\begin{equation}
X^M=(t,r,x_\perp)\,, \quad \partial_\tau X^M = (\dot{t}, \dot{r},0),\,\quad  \partial_\sigma X^M=(\frac{\Delta t}{\pi},0,\frac{b_\perp}{\pi})\,.
\end{equation}
The two D0 branes are separated in time by $\Delta t =2r_0 \eta$ resulting in a relative boost $\eta$; they are also separated by $b_\perp$ in the transverse space. The on-shell action~\eqref{eq:action} for the string becomes
\begin{equation}\label{eq:actionSchwarzschild}
S = -\frac{1}{\alpha'} \int dr \sqrt{\frac{b_\perp^2}{C^2-f(r)}+\Delta t^2} \equiv \int dr \omega(r) \,.
\end{equation}
The string oscillator level is captured by substituting $b_\perp^2 \to b_\perp^2 +\alpha' n $.
In a nice slicing of the Schwarzschild black hole the string action at oscillator level $n$ becomes
\begin{equation}
S_n =\int dT \omega_n(T)\,, \quad \text{with}\quad  \omega_n(T) = -\frac{1}{\alpha'} \sqrt{\frac{b_\perp^2+ n \alpha'}{C^2-f(r)}+\Delta t^2} \frac{dr}{dT}\,.\label{eq:niceactionSchwarzschild}
\end{equation}
We  can now compute the Bogoliubov coefficient on nice slices in Schwarzschild black holes using the methods of~\S~\ref{sec:stringproduction}.

\subsection{Real-time estimate}\label{ssec:Schwarzschildestimate}

As in Rindler space we first diagnose, using the figure of merit~\eqref{eq:betaestimate}, whether string production is adiabatic or not by computing the total number of open strings produced at the horizon~\eqref{eq:Ntotgeneral}.
We consider strings whose size is smaller than the curvature scale given by $r_0$.\footnote{We compute the size of strings along nice slices in~\S~\ref{sec:size}.} We can then use the expression $\rho(n)\sim e^{\sqrt{8\pi^2 n}}$ for the density of string states at large oscillator level yielding
\begin{equation}\label{eq:NtotSchwestimate}
 N_{tot} \approx \sum_{n\gg 1}^{\infty} e^{\sqrt{8\pi^2 n} -|\omega_n^2/\dot{\omega}_n|_{horizon}} \,,
\end{equation}
where $\omega_n$ is given by~\eqref{eq:niceactionSchwarzschild} and $\dot{}\equiv d/d T$. At the horizon $r=r_0$ the estimate is:
\begin{equation}
\left|\frac{\omega^2_n}{\dot{\omega_n}}\right|_{horizon} \hspace{-0.5cm}= \frac{4 C r_0 R_c^2 (2C^2 e r_0^2 +R_c^2)(b^2 +\alpha' n + C^2 \Delta t^2)^{3/2}/\alpha'}{(b^2\!\!+\!\! \alpha'n)(8 C^4 e^2 r_0^4 + 4 C^2 (1\!-\!C^2)e r_0^2 R_c^2\! - \!R_c^4) + C^2 \Delta t^2 (8 C^4 e r_0^2 (e r_0^2 \!-\!2 R_c^2)\!-\!3 R_c^4)}\,.
\end{equation}
In the limit of large $C$ the contribution from the high $n$ oscillator levels becomes 
\begin{equation}\label{eq:largeCestimate}
\left|\frac{\omega^2_n}{\dot{\omega_n}}\right|_{horizon} \sim \frac{\sqrt{n}}{\sqrt{\alpha'}C} \frac{R_c^2 r_0}{e r_0^2-2R_c^2}\,.
\end{equation}
Since $R_c < r_0$ (this follows from the definition of nice slices) the denominator never vanishes. 
The total number of produced open strings $N_{tot}$ is non-adiabatically enhanced for
\begin{equation}
C \gtrsim \frac{R_c^2 r_0/\sqrt{\alpha'}}{\sqrt{8\pi^2}(e r_0^2-2R_c^2)}\,.
\end{equation}
For $R_c \lesssim r_0$ string production is enhanced for $C \gtrsim \mathcal{O}(r_0/\sqrt{\alpha'})$.

In the limit of small $C$ the contribution from the high $n$ oscillator levels becomes
\begin{equation}\label{eq:largeCestimate}
 \left| \frac{\omega_n^2}{\dot{\omega}_n}\right|_{horizon} \sim 4 \sqrt{n}C \frac{r_0}{\sqrt{\alpha'}} \,.
\end{equation}
and the total number of produced open strings $N_{tot}$ is non-adiabatically enhanced for $C\lesssim\mathcal{O}(\sqrt{\alpha'}/r_0)$. Note, however, that $C<1$ corresponds to the case where the D-brane trajectories crossed (up to the impact parameter $b_\perp$) before entering the black hole.

The non-adiabaticity conditions $C \gtrsim \mathcal{O}(r_0/\sqrt{\alpha'})$ and $C\lesssim\mathcal{O}(\sqrt{\alpha'}/r_0)$ are the same conditions as in~\cite{Silverstein:2014yza} where the figure of merit at the horizon was computed in a ``natural'' slicing (in the above classification) using Gullstrand-Painlev\'{e} coordinates. Since the quantity $\omega^2/\frac{d\omega}{dT}$ is not covariant, the condition for non-adiabaticity did not have to agrees in two very different slicings of the Schwarzschild geometry. The fact that it does gives credence to the expectation of~\cite{Silverstein:2014yza} that $\omega^2/\frac{d\omega}{dT}$ is a reliable estimate of non-adiabaticity.  Note that in both, the small and the large $C$ limit, the relative boost $\eta$ between the D0 branes has dropped out.

\subsection{Saddle point method}\label{ssec:Schwarzschildsaddle}

To capture string production in the first quantized method using the saddle point approximation we need choose the relevant contour and integrate around the branch points of $\omega_T=\omega(r(T)) \frac{dr}{dT}$. Since $\frac{dr}{dT}$ is smooth in the range $r_\infty<r<r_0$ we can replace the $T$ integral by the $r$ integral:
\begin{equation}
 S=\int_{T_i}^{T_f} \omega(T) dT = \int_{r_i}^{r_f} \omega(r) dr\,.
\end{equation}
The integrand has a branch cut between an integrable square root singularity at
\begin{equation}
 r_{*,1}=\frac{r_0}{1-C^2}\,,
\end{equation}
and a branch point at at
 \begin{equation}
  r_{*,2}=\frac{r_0}{1-C^2-\frac{b_\perp^2}{\Delta t^2}}\,.
 \end{equation}
The contribution to the the action that captures string production comes from the part of the contour going around the cut between the branch point and the singularity and yields:
\begin{equation}
 S=-\frac{1}{\alpha'}\int   dr \sqrt{\frac{b_\perp^2}{C^2-f(r)}+\Delta t^2}=-i\frac{b_\perp^2 \pi r_0}{2\alpha'(C^2-1)^{3/2} \sqrt{b_\perp^2+(C^2-1) \Delta t^2}}\,.
\end{equation}
The result, in particular whether or not the action has an imaginary part, depends on the values of $\Delta t, b_\perp$ and $C$.\footnote{It is interesting to note that f or  $1-\frac{b_\perp^2}{\Delta t^2}<C^2<1$ the action has no imaginary part.} For $C>1$ or $C^2<1-\frac{b_\perp^2}{\Delta t^2}$ the Bogoliubov coefficient is
\begin{equation}\label{eq:SchwarzschildBogoliubov}
 |\beta^2|\sim e^{-\frac{b_\perp^2 \pi r_0}{\alpha' \Delta t} \frac{1}{(C^2-1)^2 \sqrt{1+\frac{b_\perp^2}{\Delta t^2}\frac{1}{(C^2-1) }}}}\,.
\end{equation}
One can include the oscillator level dependence by the replacement $ b_\perp^2 \rightarrow  b_\perp^2 +\alpha'n$.

There are two interesting limits we can consider: (1) the large boost limit $\Delta t \rightarrow \infty$ while keeping the oscillator level $n$ fixed and (2) the large oscillator level limit $n \rightarrow \infty$ keeping $\Delta t$ fixed. Note that these two limits correspond to different terms in the infinite oscillator level sum in the expression~\eqref{eq:Ntotgeneral} for the total number of produced strings.
\begin{itemize}
\item[(1)] In the large boost  limit $\Delta t \to \infty$ for fixed $n$, the Bogoliubov coefficient becomes
\begin{equation}\label{eq:Bogoliubovsaddleboost}
 |\beta^2_n|\sim e^{-\frac{(b_\perp^2 + \alpha' n)\pi}{2\alpha' \eta (C^2-1)^2}}\,.
\end{equation}
For D0 branes moving on relatively relativistically boosted trajectories in a nice slicing of the Schwarzschild geometry there is thus an enhanced production of open strings between them.\footnote{It is also important to mention that at large boost the backreaction of the open strings on the D-branes' motion becomes relevant, slowing down the motion of the D-branes \cite{McAllister:2004gd}.}
Using $R=r_0/(1-C^2)$ in the $C<1$ regime we recover the result (85) of~\cite{Silverstein:2014yza}. Our expression~\eqref{eq:SchwarzschildBogoliubov} is more general. In particular, it contains the $C>1$ regime.\footnote{In~\cite{Silverstein:2014yza} it was noted that in the $C>1$ regime the branch point and singularity, $r_{*,1}$ and $r_{*,2}$, are located on the negative real axis and it was argued that this computation would be out of the regime of control and beyond the timescale of interest for the question of near-horizon non-adiabaticity. While one may worry about the issue of regime of control for coordinates that intersect the singularity at $r=0$ such as the Gullstrand-Painlev\'{e} coordinates, we do not think that this is an issue for our nice slices which avoid the singularity. Performing the saddle point calculation then amounts to the mathematical prescription of deforming the contour in the complex plane and our result should be meaningful so long as we avoid the singularity.} 
From~\eqref{eq:Bogoliubovsaddleboost} we see that
boost-enhanced non-adiabatic string production is even further enhanced when the energy of the D-branes is larger than their rest mass, $C\gg1$. This effect is absent in flat space where the D-brane velocity drops out~\eqref{eq:flatBogoliubov}.  

\item[(2)] In the large oscillator level limit $n \rightarrow \infty$ with $\Delta t$ fixed, we get
\begin{equation}
|\beta_n^2| \sim e^{-\frac{\sqrt{n} \pi r_{0}}{\sqrt{\alpha'}(C^2-1)^{3/2}}}\,.
\end{equation}
Notice that the result scales like $\sqrt{n}$ and the $\Delta t$ dependence drops out in this limit consistent with the real-time estimate~\eqref{eq:largeCestimate}. We stress that it is this limit which is relevant for investigating a potentially non-adiabatic production rate. We again observe that $N_{tot}$ can be large (\emph{i.e.} not exponentially suppressed) for large $C$ independent of the boost $\eta=\Delta t/2r_0$.
\end{itemize}

\subsection{Real-time string production method and niceness of nice slices}\label{ssec:Schwarzschildscattering}

To have a real-time description of the string production process that goes beyond the estimate we now set up the string production problem on nice slices.

\subsubsection*{Asymptotic {\it in} and {\it out} states}
To have a well-defined string production problem we need to show that the equivalent of well-defined asymptotic \emph{in} and \emph{out} states exists for our string states. As described in~\S\ref{sec:stringproduction}, adiabatic asymptotic vacua are well-defined as long as the adiabaticity figure $\dot \omega/\omega^2$ vanishes in the far past and far future. 

\begin{figure}[ht!]
\centering{
\subfigure[The behavior of $\omega^2(T)$.]{
\includegraphics[width=.44\textwidth]{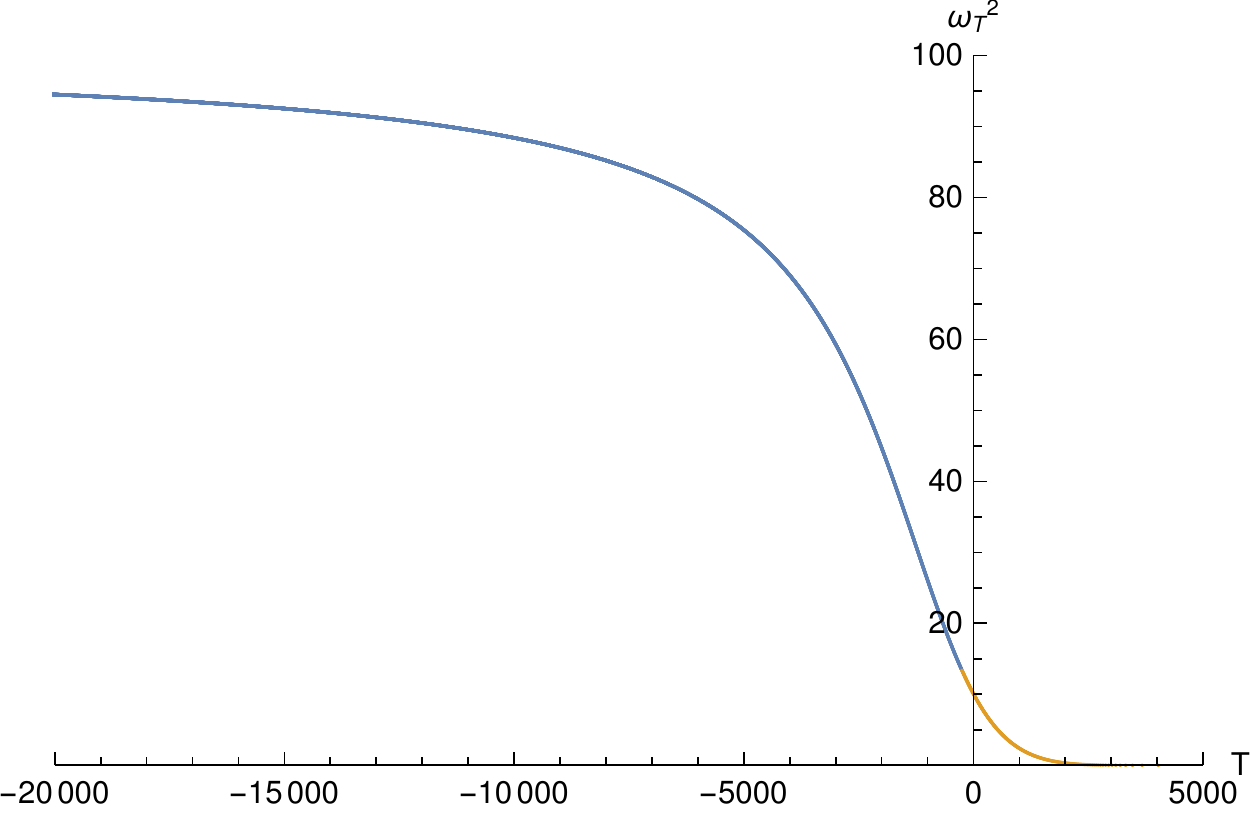}
\hspace{0.2cm}}       
\subfigure[The behavior of $dr/dT$.]{
\includegraphics[width=.44\textwidth]{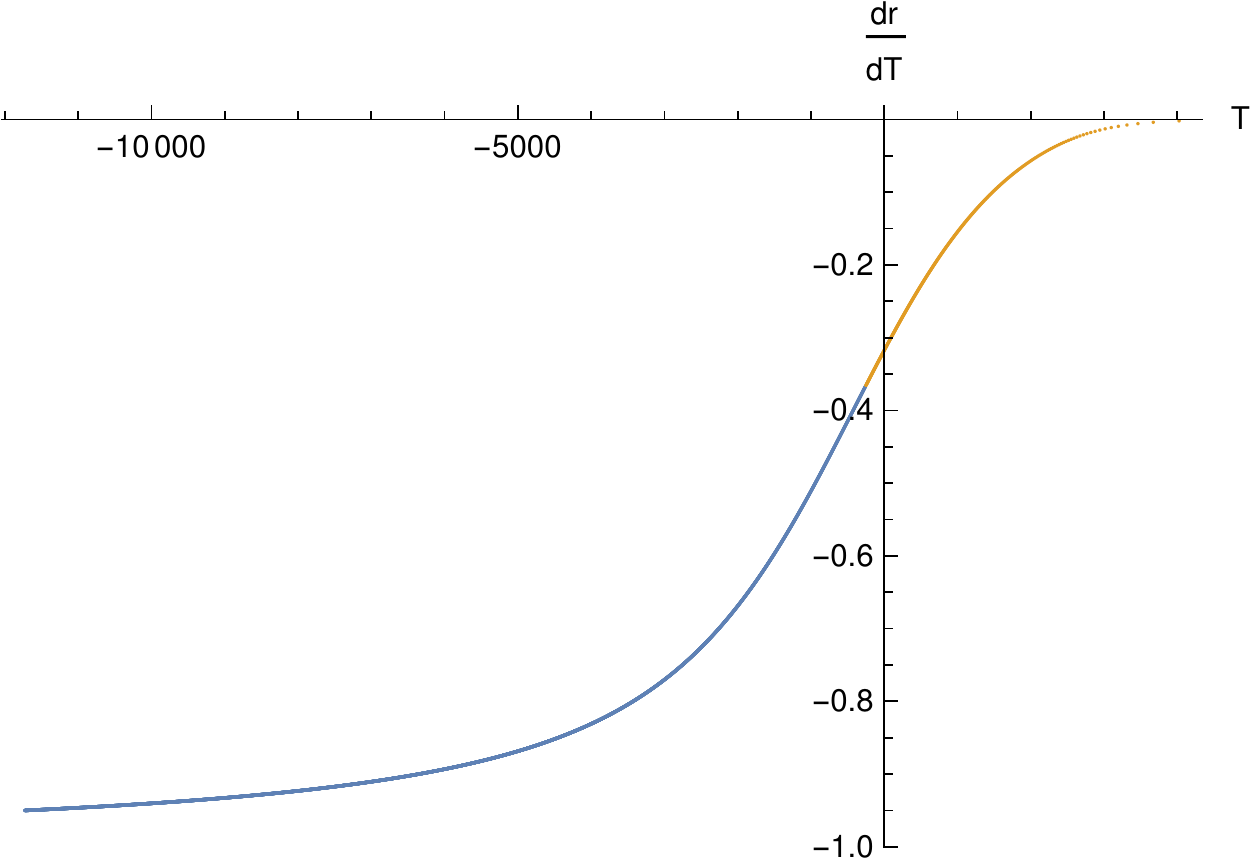}}
}
\caption{\small The behavior of (a) $\omega^2(T)$ and (b) $dr/dT$ for the choice of parameters: $\alpha'=1$, $n=10^5$, $r_0=500$, $R_c=499$, $C=300$, $\Delta t= 10$, $b=5$, $R=2 \times 10^4$. The blue segment corresponds to the region outside the horizon region while the orange segment corresponds to the region inside the horizon.
\label{fig:omegaT2}}
\end{figure}

Using \eqref{eq:Tofr} and \eqref{eq:drdtSchwarzschild} to compute the asymptotic values of $\omega_n(T)$ we note that the function $dr/dT$ goes to the constant value $\sqrt{C^2-1}/C$ for $T \to -\infty$ and approaches zero for $T\to \infty$. 
The asymptotic values of $\omega_n(T)$ in the far past, $\omega_n^{\rm in}$, and far future, $\omega_n^{\rm out}$, are then given by
\begin{equation}
 \omega_n^{\rm in} =\frac{\sqrt{b_\perp^2+\alpha' n+(C^2-1) \Delta t^2}}{C}\,, \quad \omega_n^{\rm out}  = 0\,.
\end{equation}
Therefore, we immediately see a potential problem at future infinity: while $\dot \omega_n$ vanishes for $T\to \infty$, so does $\omega^2_n$. At the end of this section we will see that this is a direct consequence of the fact that nice slices accumulate as $T \to \infty$. Hence we need to address the actual behavior of the figure $\dot \omega/\omega^2$ at late times. 
In Figure~\ref{fig:omegaT2} (a) we plot $\omega(r(T))$ which is obtained by inverting the nice slice time $T(r)$~\eqref{eq:Tofr} numerically. Its shape is determined by $dr/dT$ plotted in Figure~\ref{fig:omegaT2} (b).

\subsubsection*{Analytic model}
We just argued that having the frequency go too rapidly to zero at late times could be problematic.
We can illustrate its the implications on pair production with the following  analytic model. Consider a massive quantum field with mass $M$ in an isotropic time-dependent background. Its equation of motion can be written as
\begin{equation}
\label{exactmodel}
\ddot{\phi}_k + \Omega^2_k(T) \phi_k = 0\,,
\end{equation} 
where $\dot{}\equiv d/dT$ and $k$ denotes the magnitude of the spatial momentum.
Our nice slice frequency $\omega_n(T)$ shares similar features with this model if we take
\begin{equation}
\label{eq:toyfrequency}
\Omega^2_k(T) = k^2 + M^2(1- \tanh(\gamma T))\,.
\end{equation}
This quantum mechanical system behaves as a simple harmonic oscillator with frequencies
\begin{equation}
\Omega_k^{\rm in} = \sqrt{k^2+2M^2} \quad \text{and} \quad \Omega_k^{\rm out} = |k|\,,
\end{equation}
in the infinite past and infinite future respectively.
Note that in contrast to our problem on nice slices where $\omega_n^{\rm out}=0$, in this model we have a non-vanishing \emph{out} frequency $\Omega_k^{\rm out} = |k|>0$ instead.  The Bogoliubov coefficients $\alpha_k$ and $\beta_k$ relate the \emph{in} and \emph{out} Fock states, and if the system starts in the vacuum $|0_{\rm in}\rangle$, the mean value of particles created in the \emph{out} $k$-mode is given by 
\be
\langle 0_{\rm in} |N_k^{\rm out} |0_{\rm in} \rangle = |\beta_k|^2. 
\ee
For the system defined by \eqref{eq:toyfrequency} one can actually compute $\beta_k$ exactly\footnote{See, for instance, section 3.4 in~\cite{Birrell:1982ix}.} yielding
\begin{equation}
\label{eq:betaexact}
|\beta_k|^2 = \frac{\sinh^2(\pi (\Omega_{\rm out}-\Omega_{\rm in})/2\gamma)}{\sinh(\pi \Omega_{\rm in}/\gamma) \sinh(\pi \Omega_{\rm out}/\gamma)}\,.
\end{equation}
To mimic our system in nice slices, where $\omega^{\rm out}_n=0$, we take the $|k|\to 0$ limit and we see that $|\beta_k|^2$ diverges as
\begin{equation}\label{eq:tanhbetaexactk0}
|\beta_k|^2\approx \frac{\gamma}{2\pi} \tanh\left[\frac{\pi  M}{\sqrt{2}\gamma}\right]\frac{1}{k} + \mathcal{O}(1)\,.
\end{equation}
Therefore, if we interpret $k$ as a regulator for the vanishing frequency at future infinity, removing the regulator yields an infinite number of produced particles!\footnote{Note, however, that the total number of particles produced is still finite once we integrate over all the $k$-modes:
\begin{equation}
 N_{tot} \sim \int d^dk |\beta_k|^2 \propto \int_0^{\epsilon} dk \frac{k^{d-1}}{k} + \int_\epsilon^\infty dk k^{d-1} |\beta_k|^2< \infty\qquad \text{for}\; d>2\,.
\end{equation}
}
Hence, we might argue that the result~\eqref{eq:tanhbetaexactk0} applied to our string production problem implies that effective field theory breaks down at late times on nice slices. This conclusion, however, turns out to be premature. 
In our string production problem we do not have exact expressions like~\eqref{eq:betaexact} but only WKB (\emph{i.e.} adiabatic) approximations. In our analytic model where the WKB approximation is controlled by $|\dot \Omega_k/\Omega_k^2| \ll 1$ we have
\begin{equation}
\label{eq:toyfigure}
|\dot \Omega_k/\Omega_k^2| = \frac{\gamma M^2 \, {\rm sech^2(\gamma T)}}{2\left(k^2+M^2(1-\tanh(\gamma T))\right)^{3/2}}.
\end{equation}
Thus the WKB limit is equivalent to taking $\gamma \sim 0$ keeping all other parameters fixed. The leading behavior in the WKB approximation is given by~\cite{Chung:1998bt,Gubser:2003vk}
\begin{equation}
\label{eq:betaWKBtanh}
|\beta_k|^2_{\rm WKB} \simeq e^{-\pi T_*^{\rm im} \Omega_k(T_*^{\rm re})}\,.
\end{equation}
Here $T_*= T_*^{\rm re} + i T_*^{\rm im}$ is the location in the complex plane where $\Omega_k(T)$ vanishes.\footnote{Since $\Omega(T)>0$ for all real $T$, all its zeroes are located in the complex plane. The expression~\eqref{eq:betaWKBtanh} gives essentially the same answer as the one described in the saddle point method in~\eqref{eq:beta_tunnel}-\eqref{eq:ImS}.} Solving $\Omega_k(T_*)=0$ we obtain
\begin{equation}
T_* = T_*^{\rm re}+  T_*^{\rm im} = \frac{1}{2\gamma} \left[\log\left(\frac{k^2/M^2 +2}{k^2/M^2}\right) \pm i \pi \right]\,.
\end{equation} 
yielding
\begin{equation}
|\beta|^2_{\rm WKB} \simeq e^{-\frac{\pi^2 |k|}{2\gamma} \sqrt{2-\frac{k^2}{k^2+M^2}}}\,.
\end{equation}
The WKB result for the Bogoliubov coefficient is exponentially small in the adiabatic regime defined by $\gamma \sim 0$ and for $|k|>0$ as it should be. From this we immediately see that taking the $k \to 0$ limit is inconsistent with the WKB approximation. Indeed for all $\gamma$, no matter how small, taking $k=0$ will make the adiabaticity figure \eqref{eq:toyfigure} eventually become large in the future when $T \gg 1/\gamma$.

This simple analytic model shows that results based on the WKB approximation break down in the asymptotic regimes when the value of the frequency approaches zero too fast. For our string production problem this means that the \emph{out} state is strictly speaking not well-defined and we cannot compute $|\beta_n|^2$ by solving the wave equation~\eqref{eq:ho} for the nice slice frequency $\omega(T)$. This forces us to define our \emph{out} vacuum at some \emph{intermediate} time. For such purpose we use the definition of \emph{adiabatic vacuum}, which is well defined provided that $\dot\omega/\omega^2$ remains small enough.\footnote{It would be interesting to apply to our problem the universal definition of \emph{time dependent} particle number recently proposed in~\cite{Dabrowski:2016tsx}.}

\subsubsection*{Late-time non-adiabaticity and breakdown of WKB}

\begin{figure}[ht!]
 \begin{center}
 \includegraphics[width=0.7\textwidth]{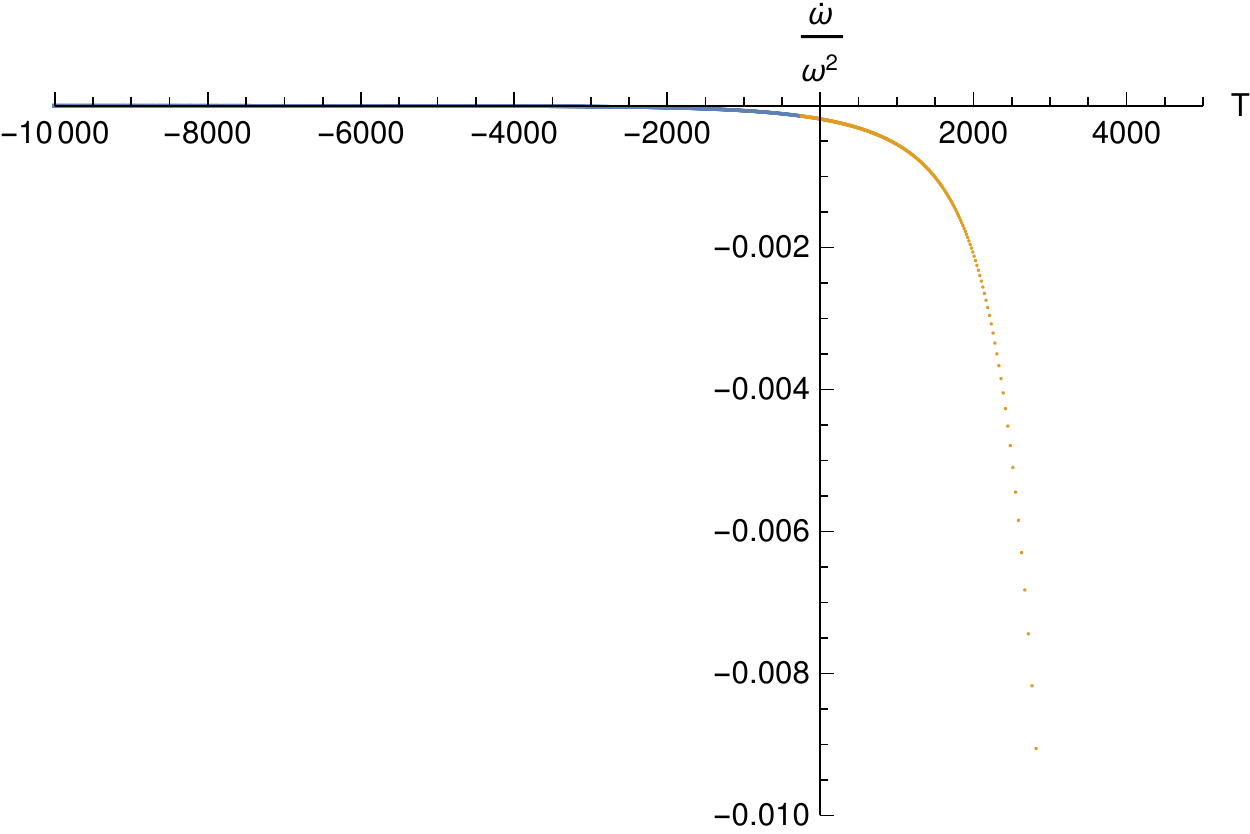}
  \caption{The blue (orange) segment corresponds to $\dot{\omega}/\omega^2$ as seen by the second infalling D0 brane from outside (inside) the horizon. Note that the D0 brane can get well inside the horizon while still being within the WKB regime ($\dot{\omega}/\omega^2 \lesssim \mathcal{O}(10^{-3})$), therefore, we still trust the WKB estimate past the horizon. The parameters used are the same ones as in Figure~\ref{fig:omegaT2}.}
  \label{fig:WKBcondition}
   \end{center}
 \end{figure}

We now investigate the physical origin of the breakdown of the WKB approximation at late times on nice slices and estimate whether there is, nevertheless, a non-adiabatic string production within the WKB regime of validity.
At late times $T \gg r_0$ the adiabaticity figure of merit at leading order grows as
\begin{equation}
\label{eq:estimate_lastslice}
\frac{\dot{\omega}_n}{\omega_n^2} \sim \frac{R_c^2 r_{\infty} e^{T/r_0}}{\frac{r_0(r_0 -r_{\infty})}{\alpha'} (X^{+})^2\sqrt{\frac{b^2+\alpha'n}{C^2-f(r_{\infty})}+\Delta t^2}} \sim  e^{T/r_0} \frac{\sqrt{\alpha'}}{r_0}\frac{C}{\sqrt{n}}\,.
\end{equation}
where in the last step we have taken the large $n$ limit. This is illustrated in Figure~\ref{fig:WKBcondition} for fixed oscillator level $n$.

For large enough
 \begin{equation}
 C \gtrsim \mathcal{O}(r_0/\sqrt{\alpha'})\,,
 \end{equation}
the figure of merit~\eqref{eq:estimate_lastslice} suggests non-adiabatic string production. Since the total number of strings produced at horizon crossing is already non-adiabatic in this limit, as suggested by~\eqref{eq:largeCestimate}, this number will continue to grow inside the black hole.

We already know that there is a limit to the validity of the WKB approximation at late times and we now discuss its origin. Recall that $\omega(T)=\omega(r) dr/dT$ with $\omega(r) \to \omega(R_c)$ going to a constant at late times. The late-time behavior of $\omega(T)$ is thus entirely due to the late-time behavior of $dr/dT$:
\begin{equation}
 \frac{dr}{dT}\sim e^{-T/r_0}\to 0\,.
\end{equation}
Hence $\omega(T)\to 0$ as $T\to \infty$. Moreover, the WKB figure 
\be
\frac{\dot \omega }{\omega^2}  \sim \frac{e^{-T/r_0}}{e^{-2T/r_0}} \sim e^{T/r_0}\,,
\ee
diverges at late times thus signaling the breakdown of the WKB approximation. But this behavior of the trajectory $r(T)$ follows from the very definition of nice slices: The nice spatial slices accumulate at late times near the last nice slice at $r_\infty$!

One may wonder whether this is a peculiar feature of the nice slices construction we are using. The accumulation effect can be traced back to the exponential $e^{-T/r_0}$ in the definition of nice slices~\eqref{eq:niceprofilesaddle}. To determine whether a different choice of $T$ dependence could avoid the accumulation effect we can replace the exponential in~\eqref{eq:niceprofilesaddle} by a general function $N(T)$ with the constraint that $N(T)\to0$ at late times $T\to \infty$ (to avoid the singularity). One finds that $\dot \omega/\omega^2 \sim \tfrac{d}{dT}(1/\dot N)$, and requiring this to vanish at late times is incompatible with $N \to 0$. Thus, we conclude that the accumulation necessarily implies the breakdown of WKB at late times irrespective of the choice of nice slice construction.

To {\it estimate} the total number of strings produced we can still use the result~\eqref{eq:estimate_lastslice} so long as we make sure that we evaluate it at times $T< T_c$ while the WKB approximation is still valid. For large enough $T_c$, there are always terms in the oscillator sum for which this is true. These terms correspond to high oscillator level $n>n_c$ for which the density of string states $\rho(n)\approx e^{\sqrt{8\pi^2 n}}$. To compute the total number of strings produced we can thus split the oscillator sum in two pieces:
\begin{equation}\label{eq:Ntotsplit}
 N_{tot}\approx \sum_{n=1}^{n_c} \rho(n) e^{-|\omega_n^2/\dot{\omega}_n|}+ \sum_{n_c}^\infty e^{\sqrt{n}(\sqrt{8\pi^2}-c)}\,,
\end{equation}
where the constant $c$ is a combination of the parameters $b_\perp, \Delta t, C, r_0$.
The first sum is not amenable to the WKB approximation. The second sum, on the other hand, is: If the constant $c\lesssim \sqrt{8\pi^2}$, then the second sum in~\eqref{eq:Ntotsplit} yields non-adiabatic string production no matter what the value of first sum in~\eqref{eq:Ntotsplit} is. Hence, if we are interested in evaluating the figure of merit at intermediate times, {\it e.g.} when the second D0 brane crosses the horizon, we can always find a regime where $\dot{\omega}/\omega^2\ll 1$ and so we can trust the WKB approximation and define an adiabatic vacuum at the horizon (as discussed \emph{e.g.} in section 3.5 of~\cite{Birrell:1982ix}).\smallskip

To extend the regime of validity of the WKB approximation to the asymptotic future we need a regulator for the vanishing asymptotic frequency $\omega_{\rm out}$. The harmonic oscillator model has such a natural regulator: the spatial momentum $k$ of the particle. What would provide such a the regulator in the string production problem on nice slices? The frequency $\omega(T)\to 0$ at late times because the nice slices accumulate $dr/dT \sim e^{-T/r_0} \to 0$. To compensate for this exponential factor we could attempt to promote the string tension $T_s=1/2\pi \alpha'$ to a function of time\footnote{We would like to thank Eva Silverstein for this suggestion.} so that $T_s \sim e^{T/r_0}$ at late times. This could be achieved in a D-brane compactification where the string descends from a D-brane wrapping an internal cycle that shrinks at late times. In that case, however, the mass of the infalling D0 branes
\begin{equation}
 m=\frac{1}{g_s \sqrt{\alpha'}}\sim \frac{e^{T/2r_0}}{g_s}\,,
\end{equation}
grows exponentially at late times implying a large backreaction effect.
A logical conclusion of these considerations might be that the late-time breakdown of the WKB approximation is inevitable and one therefore cannot set up the string production problem within this approximation. It would be interesting to study this issue further.

\section{Size of open strings}\label{sec:size}

In order to justify our non-adiabaticity analysis using the Hagedorn density of string states for large oscillator level, namely the expression $\rho(n)\sim e^{\sqrt{8\pi^2 n}}$, in computing~\eqref{eq:NtotSchwestimate} we need to make sure that the spatial extent of the produced open strings is small compared to the horizon scale. In this section we calculate the typical size of a string stretched between the pair of D0 branes along the nice slice defined by the fixed nice slice time $T_*$. For the sake of simplicity, we study this problem in Rindler space. 

\begin{figure}[ht!]
\begin{center}
 \includegraphics[width=0.5\textwidth]{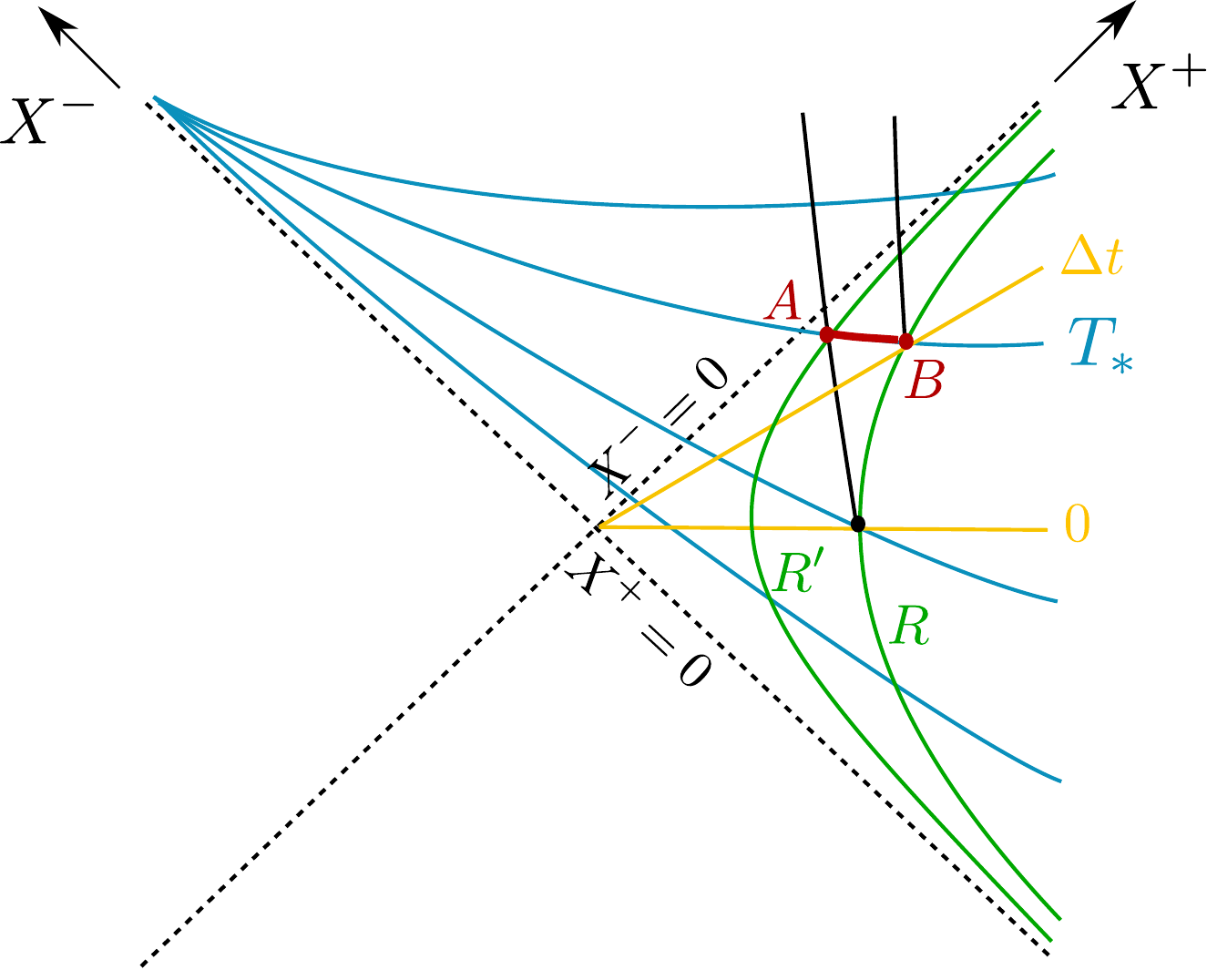}
 \caption{The blue, yellow and green lines correspond, respectively, to slices of constant nice slice time $T$, Schwarzschild time $t$ and radial coordinate $r$. The trajectories of the two D-branes are shown in black with the red segment representing the open strings stretched between the D-branes at fixed nice slice time $T_*$. }
 \label{fig:sizeofstring}
 \end{center}
\end{figure}

Using the line element of the Rindler space on the nice slice, the size $L$ of the produced open strings on a fixed nice time slice $T_*$ is computed from
\begin{equation}\label{eq:stringsize}
L= \int^{X^{+}_{B}}_{X^{+}_{A}} \sqrt{e^{-\frac{T_{*}}{r_{0}}} +\frac{R_{c}^2}{(X^{+})^2} } \;dX^{+}\,,
\end{equation}
with the end points,
\begin{equation}\label{eq:D0bdy}
X^{+}_{A} =R' e^{\frac{t(R')}{2r_{0}}}, \qquad X^{+}_{B} =R e^{\frac{\Delta t}{2r_{0}}}\,.
\end{equation}

Let us consider the case where the first D0 brane is dropped from the radial location $R$ at Schwarzschild time $t=0$, while the second D0 brane remains at $R$; the temporal separation between the branes then defines the boost. Let us denote the coordinates of the first and second D0 brane by $(t(R'), R')$,  $(\Delta t, R)$, respectively.\footnote{$t(\rho)$ is the trajectory of the free falling particle in the Rindler space.} Since these two D0 branes are on the same nice slice, we have
\begin{equation}
T_{*}=T(\Delta t, R)=T( t(R'),R'), \label{eq:stringnice}
\end{equation}
where $T(t, \rho)$ is given by 
\begin{equation}
T(t,\rho)= r_{0} \log \left[ \frac{e^{t(\rho)/r_{0}}}{\left(\frac{R_{c}}{\rho}\right)^2+1} \right]\,.
\end{equation}
We can solve (\ref{eq:stringnice}) in the limit $C \rightarrow \infty$ for the location of the first D0 brane:
\begin{equation}\label{eq:Deltateq}
R'^2 = (R^2 +R_{c}^2) e^{-\frac{\Delta t}{r_{0}}}-R_{c}^2\,.
\end{equation}

\begin{figure}[ht!]
\begin{center}
 \includegraphics[width=0.45\textwidth]{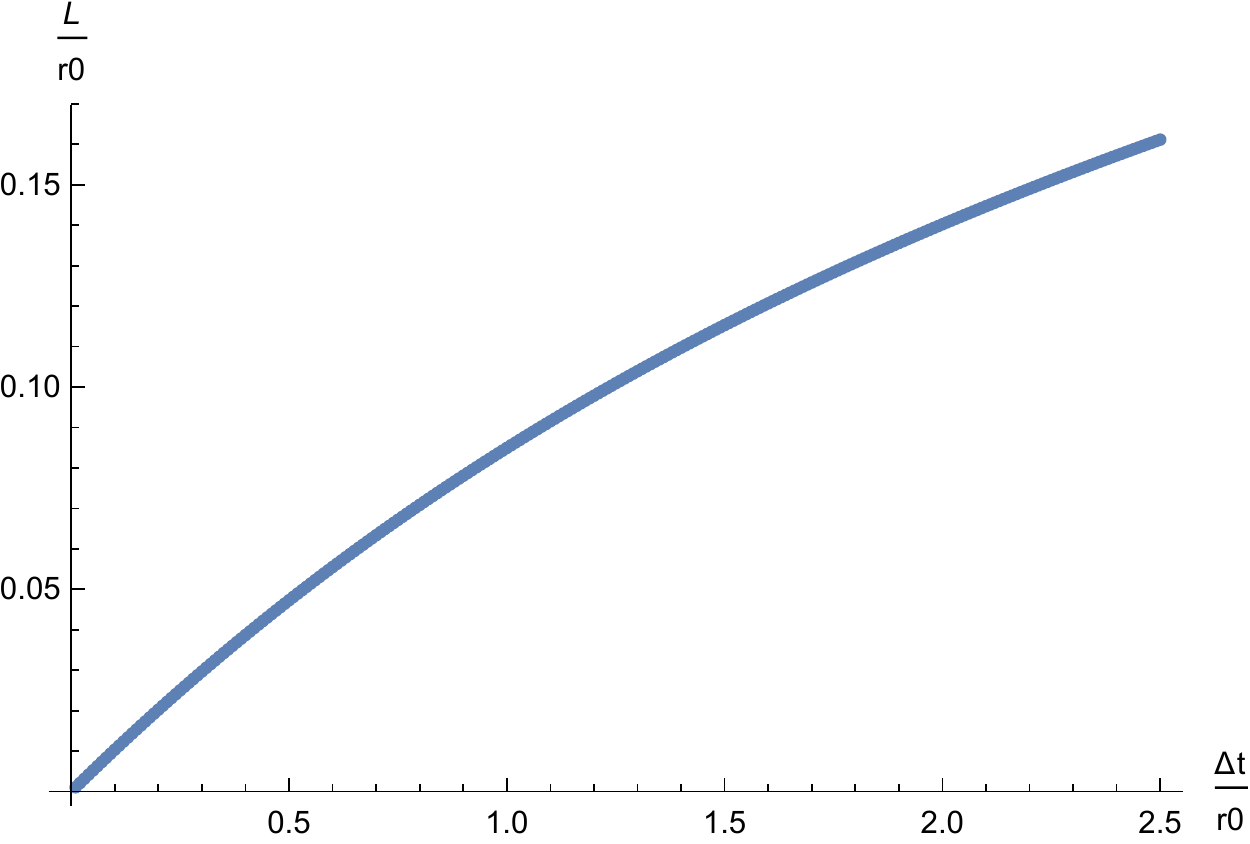}
 \caption{Typical size of a string stretched between two D0 branes for the parameters $R = 2$, $r_0 = 10$, $R_c = 1/2$ and $C = 1000$.}
 \label{fig:stringsize}
 \end{center}
\end{figure}

Now that we have specified the positions of both D0 branes~\eqref{eq:D0bdy} on the nice slice defined by $T_*$ we can evaluate~\eqref{eq:stringsize}. In figure~\ref{fig:stringsize} we plot the dependence of the string size $L$ on the boost $\Delta t$, both normalized by $r_0$.

From~\eqref{eq:Deltateq} we see that so long as the boost satisfies
\begin{equation}\label{eq:Deltatbound}
 \Delta t\leq r_0 {\rm log}\left[1+\left(\frac{R}{R_c}\right)^2\right]\,,
\end{equation}
the size of the stretched strings remains smaller than the horizon scale and, hence, the use of the string density $\rho(n)\sim e^{\sqrt{8\pi^2 n}}$ in computing the string production rate is justified. 

Recall that here we have restricted ourselves to the near-horizon region for ease of calculation. This is a good approximation for estimating the string size in the full black hole geometry provided that the separation of its end points in the Rindler coordinate is within the range $0\leq \rho \leq 2r_0 C$. By choosing $C$ large enough we can drop the D-branes from larger and larger values of $R$. Equation~\ref{eq:Deltatbound} then implies that we can make the boost relativistically large, and thus have a non-adiabatic enhancement in~\eqref{eq:Bogoliubovsaddleboost}, while keeping the string size small. This shows that our non-adiabaticity analysis is consistent.

\section{Discussion}\label{sec:discussion}

We now discuss the interpretation of our results and their implications for effective field theory on nice slices.

The notion of nice slices was originally introduced in~\cite{Lowe:1995ac,Polchinski:1995ta} in the context of the information paradox as a way to avoid the discussion of trans-Planckian physics. To derive Hawking radiation in a Hamiltonian framework entirely within low-energy field theory the spacelike slices, on which the state is defined and pushed forward, have to be sufficiently well-behaved. Assuming that low energy field theory knows its own range of validity there seems to be no room for stringy effects. Indeed there are no large local invariants in the classical black hole geometry. On the other hand, the recent firewall discussion (see the original AMPS paper~\cite{Almheiri:2012rt} and the references who cite them) and the preceding work by Mathur~\cite{Mathur:2009hf} have taught us that local effective field theory with a smooth horizon and unitarity of black hole evaporation are incompatible. To avoid information loss, at least its concise version using the strong subadditivity theorem, one may 
either give up the smooth horizon and look for a string theory realization of the new degrees of freedom at the horizon-scale or invoke non-local effects.\footnote{Looking for horizon-scale structure is being pursued in the fuzzball programme while non-local effects are invoked to different degrees in various proposals including ~\cite{Papadodimas:2012aq,Maldacena:2013xja,Giddings:2012gc}. The proposal of~\cite{Dodelson:2015uoa,Dodelson:2015toa} may fit somewhere in between.}
 
In this work we followed a more conservative route, based on the observations in~\cite{Silverstein:2014yza}, and studied stringy-theoretic effects on nice slices in the classical background of a Schwarzschild black hole. A one-loop string process capturing open string pair production is found to give rise to non-adiabatically enhanced effects. For some regime of parameters of the infalling D-brane system we find that the figure of merit~\eqref{eq:largeCestimate} and the saddle point result~\eqref{eq:Bogoliubovsaddleboost} are enhanced, essentially due to the Hagedorn density of string states.  At late nice slice time we find that due to an accumulation of slices the system becomes non-adiabatic. These effects suggests a breakdown of effective field theory on nice slices catalyzed by the late-time infalling observer. We now interpret our results and discuss their potential implications. Finally we will conclude our discussion by commenting on the relation to other works 
in the literature.

\paragraph{(\emph{i})  Non-adiabaticity from Hagedorn density of string states.}

In the body of the paper, we employed the strongest form of non-adiabaticity, namely, a large total number $N_{tot}=\sum_n \rho(n) |\beta_n|^2$ of open strings created. This requires that the figure of merit $\omega_n^2/\dot{\omega}_n$ grows slower than logarithm of  the density of string states $\rho(n)$. We found that this is possible when the energy $E$ of the two infalling D-branes is large compared to their rest mass $m$, with their ratio being typically of order $C \equiv E/m\gtrsim \mathcal{O}(r_0/\sqrt{\alpha'})$. This is in agreement with~\cite{Silverstein:2014yza}. For ultrarelativistic D-branes the resulting geometry is a gravitational shockwave which may explain the enhanced string production for $C\gg 1$. We also find non-adiabatic production for $C\lesssim\mathcal{O}(\sqrt{\alpha'}/r_0)$ but recall again that for $C<1$ the D-brane trajectories crossed before entering the black hole.

Note that for both, the large and the small $C$ limit, the relative boost $\eta\equiv \Delta t/2r_0$ between the D-branes drops out. This is rather curious as the large boost which naturally arises in the near-horizon region of black holes was the original motivation for setting up the Gedankenexperiment - with the expectation that string creation is much more enhanced than the na\"{\i}ve extrapolation from local effective field theory would suggest.

While our results may to some extent be interpreted as a breakdown of effective field theory on nice slices near the horizon of black holes, they may not immediately be connected to the firewall. One reason is that the onset of non-adiabaticity is not confined to the near-horizon region. As we showed in~\S~\ref{ssec:flatestimate} string production can be enhanced farther outside the horizon with the onset of non-adiabaticity determined by the parameters of the system. Perhaps unsurprisingly, string production gets further enhanced inside the horizon. Another reason is that, in the way the Gedankenexperiment is set up, a potentially non-adiabatic effect as seen by the late infalling observer (modeled by the second D-brane) depends on having an earlier infalling system (the first D-brane). Moreover, the non-adiabatic enhancement of open string pair production only seems to occur in extreme kinematic regimes and is thus not a feature of generic choices of parameters. One may get around these restrictions in 
other 
systems but it remains to 
be seen whether string-theoretic effects like the one considered here can provide a generic dynamical mechanism for a breakdown of effective field theory.

Besides the strong version of non-adiabaticity just discussed there is the possibility of a weaker version. For example, one can argue that there is significant string production if the total mass of the produced strings is larger than the black hole mass:
\begin{equation}
M_{tot}^2 = \sum_n^{\infty} n e^{\sqrt{8\pi^2 n} -|\omega_n^2/\dot{\omega}_n|} \sim M_{BH}^2. \label{eq:weaknonad}
\end{equation}
In that case we need to take backreaction effects serious and the analysis based on a fixed background breaks down. It is not clear to us whether this type of non-adiabaticity is genuine stringy physics or can be captured by effective field theory. In any case it would be an interesting future problem to elaborate this possibility further.

\paragraph{(\emph{ii}) Non-adiabaticity from accumulation of nice slices.}

In a nice slicing of the black hole geometry we expect that the use of effective field theory is justified because it avoids the black hole singularity and, generally, regions of large spacetime curvature. Our results, however, imply that this is not the case. In~\S~\ref{ssec:Schwarzschildscattering} we found that at late times string production becomes non-adiabatic and eventually the WKB approximation breaks down. We explained that this breakdown is due to an accumulation of nice slices near the last nice spatial slice. 

The appearance of highly excited states due to the accumulation is understood by the usual uncertainty argument. Let $E_{t}$, and $\Delta t$ be the energy and time displacement measured in Schwarzschild coordinates and $E_{T}$ and $\Delta{T}$ be those in nice slice coordinates. Since $\Delta t =f(T)\Delta T$, where $f(T)\rightarrow 0$ at late time $T \rightarrow \infty$, a typical excitation on the nice slice becomes lighter eventually {\it i.e.}, $E_{T}=f(T) E_{t} \rightarrow 0$ even though the energy in the Schwarzschild coordinate $E_{t}$ is originally very large. In particular, heavy stringy modes become light in the late time limit, hence the production rate is non adiabatically enhanced.  

We suspect that this kind of breakdown of effective field theory not only happens in our particular construction, but is a generic feature of nice slices. This is because in order to avoid the black hole singularity the slices must accumulate somewhere in the spacetime, therefore $\omega^2/\dot{ \omega} = 0$ at the last nice slice. The place where the accumulation occurs depends on the particular construction and so the on-set of non-adiabaticity and WKB breakdown is coordinate dependent. Or course, we are not claiming that accumulation has locally detectable physical consequences. 
Indeed one may avoid accumulation by going to a different (non-nice) slicing. However, in that case one has to face the singularity that one inevitably encounters. Our observation here merely is that even though nice slices are most suited to describe the black hole evolution, effective field theory inevitably breaks down due to the accumulation of slices regardless of choice of nice slicing. One may be tempted to conclude that to be able to fully describe the evaporation process and to address the information loss problem we need to better understand the relevant aspects of full fledged quantum gravity.

\paragraph{(\emph{iii}) Relation to other works.}

Our result that the highly excited string states modify the low energy effective description on nice slices bears a resemblance to other works in the literature.

In~\cite{Lowe:1995ac,Polchinski:1995ta} it was argued that the relative Lorentz boost of the infalling body and asymptotic observer gives rise to large non-local invariants which could invalidate low energy effective field theory. Commutators of low energy fields in string field theory on nice slices were found to be non-vanishing even when the operators were spacelike separated - contrary to the behavior of the commutators in local field theories. From this the authors argued that the correct effective field theory in the black hole geometry must contain highly non-local states of extremely massive strings stretched between distant points on the nice slice. The effects of these stringy modes becomes especially significant in the late time limit where the mass squared becomes trans-Planckian while the nice-slice energy remains low. While such an effect would invalidate the use of low energy field theory on nice slices it has, to the best of our knowledge never been fully settled whether this is a real 
effect or a gauge artifact. However, their picture is consistent with the results of this paper that highly excited string states should be included in the low energy effective description derived from string theory. We found that the figure of merit grows exponentially in time for every string mode, therefore the appearance of these highly excited strings at late time in the effective theory is unavoidable. It would be interesting if one could make a more detailed connection between these two analyses.
 
Another interesting string-theoretic effect that is enhanced near the horizon of black holes is string spreading. It has long been known that strings have an unusual property at large boost: transverse growth~\cite{Susskind:1993ki,Susskind:1993aa}. To address the information paradox, however, a non-local effect is needed along the nice slice, in the longitudinal direction. Recently, it has been argued~\cite{Dodelson:2015toa} that a detectable level of root-mean-square longitudinal spreading of an early infalling string could be measured by a late infalling detector.\footnote{See~\cite{Dodelson:2015uoa} for an exhibition of longitudinal non-locality in well-defined gauge-invariant S-matrix calculations.} An interesting observation arising from kinematic considerations is that in order to observe a breakdown of effective field theory one has to consider processes involving secondary probes, {\it i.e.} a third system in addition to the early and late infallers (such as {\it e.g.} a string emitted in the near-
horizon region by the late infaller). It is tempting to look for a connection between this result and our findings that non-adiabatically enhanced production of open strings on nice slices between an early and a late infalling D-brane appears only in extreme kinematic regimes. We leave this exciting possibility for future work.

 \smallskip

\section*{Acknowledgements}
We are especially grateful to Joe Polchinski and Eva Silverstein for extensive and insightful discussions on several aspects of this work.
We would like to thank Matthew Dodelson, Daniel Harlow, Gary Horowitz, Don Marolf, Samir Mathur, Edgar Shaghoulian, Mark Srednicki and Sasha Zhiboedov for many useful discussions. We are particularly thankful to Samir Mathur, Joe Polchinski and Eva Silverstein for comments on the manuscript. We have also benefitted from discussions with the participants of the KITP Program: Quantum Gravity Foundations: UV to IR.
The work of A.P. is supported by National Science Foundation Grant No. PHY12-05500. A.P. acknowledges support from the Black Hole Initiative (BHI) at Harvard University, which is funded by a grant from the John Templeton Foundation and support from the Simons Investigator Award 291811 and the DOE grant DE-SC0007870.
The work of F.R. is supported in part by FAPESP grants 2012/05451-8 and 2014/17363-1 and the National Science Foundation under Grant No. NSF PHY-1125915. T.U is supported by National Science Foundation under Grant No. NSF PHY-25915. F.R. would like to thank the DFI-FCFM Universidad de Chile for hospitality of during the last stages of this work.

\bibliographystyle{toine}
\bibliography{references.bib}

\end{document}